\documentclass[twocolumn,showpacs,aps,amssymb,floatfix,prd,amsmath,preprintnumbers]{revtex4} 
\usepackage{graphicx}  
\usepackage{dcolumn}   
\usepackage{bm}
\begin{document}

\def \reals{{\mathbb R}}
\def\be{\begin{equation}}
\def\ee{\end{equation}}
\def\bea{\begin{eqnarray}}
\def\eea{\end{eqnarray}}
\def\nn{\nonumber}
\def\th{\theta}
\def\ph{\phi}
\def\lt{\left}
\def\rt{\right}
\def\ed{\end{document}}
\def\degree{\mathop{\rm {{}^\circ}}}
\input epsf.tex

\title{Relativistic images of Schwarzschild black hole lensing}\thanks{This work is dedicated to the memory
of Professor John ~A.~Wheeler who coined the term {\em black hole.}}

\author{K.~S.~Virbhadra}
    \email[]{shwetket@yahoo.com}
     \affiliation{Department of Mathematics, Physics and Statistics, University of the Sciences in Philadelphia, Philadelphia, Pennsylvania 19104,USA}

\begin{abstract}
We model massive dark objects at centers of many galaxies as Schwarzschild black hole lenses and study
gravitational lensing by them in detail. We show that the ratio of mass of a Schwarzschild lens to the 
differential time delay between outermost two relativistic images (both of them either on the primary
or on the secondary image side) is  extremely insensitive to changes in the angular source position as 
well as the lens-source and lens-observer distances. Therefore, this ratio can be used to obtain very 
accurate  values for masses of black holes at centers of galaxies. Similarly, angular separations between
any two relativistic images are also extremely insensitive to changes in the angular source position 
and the lens-source distance. Therefore, with the known value of mass of a black hole, angular separation
between two relativistic images would give a very accurate result for the distance of the black hole. 
Accuracies in determination of masses and distances of black holes would however depend on accuracies 
in measurements of differential time delays and angular separations between images. Deflection angles 
of primary and secondary images  as well as effective deflection angles of  relativistic images on the
secondary image side are  always positive. However, the effective deflection angles of relativistic 
images on the primary image side may be positive, zero, or negative depending on the value of angular 
source position and the ratio of mass of the lens to its distance. We show that effective deflection 
angles of relativistic images play significant role in analyzing  and understanding strong gravitational 
field lensing.
\end{abstract}



\maketitle

\section{\label{sec:Intro}Introduction}

Light deflection in weak gravitational field  of Schwarzschild spacetime  is well-known  since 
1919 \cite{WkSt,Edd20}, and it serves as the  starting point to the learning of  gravitational  lensing (GL) theory even now 
\cite{book1,book2,NB01Wu07}.  However, light deflection in strong gravitational field of  Schwarzschild 
spacetime was not studied until around 5 decades ago Darwin \cite{Dar59and61} pioneered a theoretical 
research  on  GL resulting from   large deflection of light in the
vicinity of  photon sphere of Schwarzschild spacetime. A few years later, Atkinson \cite{Atk65} 
extended these studies to a general static spherically symmetric spacetime. In fact, apart from a few activities, 
the subject of strong gravitational field lensing remained in almost a dormant stage
until toward the end of the last century, and this was due to two  main  reasons. First, Darwin's calculations 
showed that the images are very  demagnified and therefore those are very difficult to be observed with the available 
observational facilities. Secondly, the known gravitational lens equation (see in \cite{book1,book2}) was not 
adequate for the study of lensing due to large deflection of light. As  astronomical 
techniques are improving fast, it may be possible  to  overcome the observational  obstacles in future. 
Therefore, an adequate lens equation   was required for  this purpose. To this end, Frittelli and 
Newman \cite{FN99} obtained an exact lens equation that is  applicable to  arbitrary spacetimes;
however, this equation is difficult to use in general. Further, under some physically realistic assumptions,
we \cite{VE00} obtained a simple lens equation that allows arbitrary small as well as large light deflection angles.
Later, Frittelli {\it et al.} and   Kling {\it et al.} \cite{FritEtalKlingEtal00}  carried out 
comprehensive comparative studies  of the  exact lens equation  with our lens equation for the case of
Schwarzschild spacetime. They found that our lens equation works remarkably well 
as both approaches  to  gravitational lensing  yield  extremely close results even for light 
rays which have large deflections in strong gravitational field and  go around the lens several times before 
reaching the observer. As our lens equation is easy to use and yields very  close results to those obtained 
by using an exact lens equation, our lens equation
has been most widely used in the literature for studying strong field gravitational lensing. 
Perlick \cite{Per07}, in a recent brief review, called our lens equation an {\em almost exact lens equation}.
In last 8 years, there has been a  growing  interest in studying weak as well as strong field lensing  by
black holes, naked singularities, wormholes, and some other exotic objects 
(see \cite{PerEtal,Amoreetal,ZakEtal,many1,many2,Bozetal01,BozMan04,VNC98,VE02,VK08,NSGL,WH,Others} and references therein).

In \cite{VE00}, we modeled the massive dark object (MDO) at the Galactic center as a Schwarzschild black hole lens, 
and, using our lens equation, studied point source GL due to light deflections in weak as well as strong
gravitational fields. By solving the lens equation numerically, we obtained image positions and 
 their magnifications. Like in Darwin's paper, our computations showed that, in addition to the weak field primary 
(also called direct) and secondary images, there are  theoretically an infinite sequence of very  demagnified images
on both sides of the optical axis; we named them {\em relativistic 
images}. With increase in the value of angular source position (measured with respect to the optical axis),
magnifications of relativistic images decrease with much faster rates compared to magnifications of 
primary and secondary images. Therefore, relativistic images are not just very demagnified, but are also 
transient. However, despite several other  observational difficulties (discussed in \cite{VE00} and also in the 
last section of this paper), if these images
were ever observed, we showed that it would give an upper bound  to the compactness of the MDO.
Therefore, it would push  black hole interpretation of the MDO at the Galactic center. Observation of
relativistic images would  be undoubtedly a landmark discovery in astronomy; however, experimentalists and
astronomers have to pay the price for these very important observations.  When the source, the lens, and
the observer are aligned, in addition to an Einstein ring, we get theoretically an infinite sequence of rings;
we \cite{VE00} termed them {\em relativistic Einstein rings}.

The central thread in this paper is a comprehensive study of relativistic images of Schwarzschild black hole
lensing. The purpose of this  is to theoretically investigate if  possible observations of these images, 
compared to primary and secondary images, could provide  more valuable and accurate   information about the lens.
Though the primary and secondary images of Schwarzschild black hole lensing are well discussed in the literature, we 
also include these in this paper for 3 reasons. First, we do  not take either weak or strong field approximations in our
computations. Thus,  our results are more accurate than  known  weak field limit approximate  analytical 
expressions could provide for lensing observables for these images. Therefore, these more accurate results
could be useful for observations in near future with advancing astronomical facilities. Secondly, it is useful 
to have  thorough  studies of relativistic as well as primary and secondary images due to a  gravitational lens (with the same mass
and distance) in the same
paper so that one can immediately compare properties of these images. For observations of relativistic images and
measurements associated with them, a detailed information about their primary and secondary images are helpful. 
Thirdly, differential time delay of secondary image with respect to the primary image is though well discussed
in the literature, studies of their individual time delays were not paid enough attention for some reasons (discussed
later in this paper). Therefore, we study these and obtain some important results.

This paper is organized as follows.  In Section II, we first discuss  a lens equation applicable to weak as well
as strong gravitational field regions, and then give a brief review of deflection angles and time delays of light rays
traveling in Schwarzschild spacetime. In Section III, we show that a schematic diagram for  effective deflection angles of relativistic
images give deep insight and provide some important information before computations. In Section IV, we  model our
Galactic MDO as the Schwarzschild black hole lens and study  variations in image positions,  magnifications, and
time delays of primary and secondary as well as relativistic images with changes in angular source position and
lens to  source distance. We also study the variation of deflection angles for primary and secondary images
and effective deflection angles of relativistic images with respect to changes in the angular source position and
lens to source distance. Computations of  effective deflection angles  provide 
geometrical beauty of strong  gravitational field lensing and an analysis of those support the reasonableness of
other results obtained through numerical computations. In Section V, we model MDOs at centers of many galaxies
as Schwarzschild black hole lenses and study the variations in the same physical quantities (as studied in  Section IV) with
respect to changes in the ratio of mass of lens to its distance  and the lens-source distance for a fixed value  of angular 
source position. In Section VI, we discuss and summarize the results.

Bozza {\it et al}.\cite{Bozetal01} obtained approximate analytical expressions for  image positions and 
magnifications of  relativistic images. Bozza and Mancini \cite{BozMan04} further derived approximate analytical 
expressions for differential time delays among relativistic 
images. As the aim of our paper is to present accurate results and to derive some important astrophysical implications for 
those, we do not digress to  compare and contrast our results with  approximate analytical results given in 
\cite{Bozetal01,BozMan04}. In Section IV, we briefly discuss that there are small (but significant)
percentage differences in results for image positions. However, percentage differences in results for magnifications of images
are very large. In Section V, we compare their  approximate results with ours for differential 
time delays between two outermost 
relativistic images. We  show  that there are qualitative as well as  large  quantitative   differences. We justify our results
with some arguments.

There are some  fascinating  results in this paper. The most important among those is that  relativistic images would
provide very accurate values for masses and distances of MDOs at centers of galaxies.
The ratio of the mass of a Schwarzschild lens to the differential time delays  between two outermost  relativistic images 
(both of them either on the primary image side or on the secondary image side from the optical axis) 
is almost a constant; i.e., this ratio is  {\em
extremely} insensitive to  changes in the angular source position, the observer-lens distance, and the lens-source
distance. Therefore, observation of relativistic 
images and measurements of their differential time delays would give very accurate values for  masses of  MDOs.
Another very useful property of relativistic images is that variations in their angular separations due to changes
in the angular source position and the lens-source distance are extremely small. Therefore, once we have accurate values
for masses of MDOs, measurements of angular separations between  relativistic images would give very accurate results for
distances of those MDOs. Similarly, we also show that the measurement of ratio of fluxes of outermost relativistic images
(one on each side of the optical axis) would help us obtain very accurate result for  distance of the source.

As in our previous papers on GL \cite{VE00,VNC98,VE02,VK08},
we use geometrized units (i.e., the gravitational constant 
$G=1$ and the speed of light in vacuum  $c=1$, so that $M \equiv M G/c^2$) throughout this paper. However, we   finally present   time delays and
differential time delays of images  in the  unit  of {\em minute}.


\section{\label{sec:LE&DE} Lens Equation,  Deflection Angle, and Time Delay}

Assuming that the   angular  position of source of light  is small and  the  source as well as the  observer are
situated at    large distances from the lens (deflector), we \cite{VE00} obtained a new gravitational 
lens equation that allows for arbitrary  large as well as small deflections of light. (The first assumption does not
hurt applicability of the lens equation, as GL is usually meaningful only for small angular source positions.)
The lens equation  is given by

\begin{equation}
  \tan\beta =  \tan\theta -  {\cal D} \lt[\tan\theta + \tan\lt(\hat{\alpha} - \theta\rt)\rt] \text{,} 
  \label{GravLensEqn}
\end{equation}
with
\begin{equation}
  {\cal D} =
    \frac{D_{ds}}{D_s}   \text{.}
   \label{calD}
\end{equation}

Angular positions of an unlensed source and images  are measured from the optical axis (the line joining the observer and the 
center of mass of the gravitational lens), and are 
represented by symbols  $\beta$ and $\theta$, respectively. These angles when  measured in clockwise and anticlockwise directions
from the optical axis are assigned positive and negative signs, respectively.
$\hat{\alpha}$ represents the total angle by which
the light ray is deflected in the gravitational field of the lens while traveling from the source to the observer.
Null geodesics which are bent toward and away from the lens have, respectively, $\hat{\alpha}>0$ and $\hat{\alpha}<0$.
$D_{d}$, $D_{ds}$, and $D_s$ stand, respectively, for   observer-lens, lens-source, 
and observer-source  angular diameter  distances. The values of   parameter ${\cal D}$ mathematically lie in the 
interval $\left(0,1\right)$; however, for the lens equation to hold good, the value of ${\cal D}$ should not be taken 
too close to $0$.
The perpendicular distance from the center of mass of the lens to the tangent to 
the null geodesic at the source position  is (see Fig. 1 in \cite{VE00})
\begin{equation}
J = D_d \ \sin\theta   
  \label{J}
\end{equation}
and is called impact parameter.

The magnification of an image formed due to GL is defined as the ratio of the flux of the image to the 
flux of the unlensed source. However, according to  Liouville's theorem, the surface brightness 
is preserved in GL. Therefore, the magnification $\mu$ of an image formed
due to gravitational lensing  turns out to be the ratio of the solid angles of
the image and of the  unlensed source made at the observer \cite{book1,book2,NB01Wu07}.  Thus, for a circularly symmetric 
GL, the  magnification $\mu$ of an  image is obviously expressed by
\begin{equation}
  \mu = \mu_t \mu_r \text{,}
      \label{Mu}
\end{equation}
where the tangential magnification $\mu_t$ and the radial magnification $\mu_r$ are, respectively, expressed by
\begin{equation}
  \mu_t = \lt(\frac{\sin{\beta}}{\sin{\theta}}\rt)^{-1} ~~~~~\text{and} ~~~~~~
  \mu_r = \lt(\frac{d\beta}{d\theta}\rt)^{-1}  \text{.}
        \label{MutMur}
\end{equation} 
{\em Tangential  critical curves} (TCCs) and  {\em radial critical curves}  (RCCs) are, respectively, given
by singularities in $\mu_t$ and  $\mu_r$ in the lens plane. However, their corresponding values
in the source plane are, respectively,  termed {\em  tangential caustic} (TC) and  {\em radial caustics}
(RCs). The parity of an image is called positive if $\mu>0$ and negative if $\mu<0$. Sometimes terms
magnifications and absolute magnifications of negative parity images are used synonymously. If the angular source
position $\beta=0$ (i.e., when the source, the lens, and the observer are aligned), there may be ring shaped 
image(s) [called Einstein ring(s)]; these images are assigned $0-$ parity. Note that $\beta=0$ does not always
give Einstein ring(s) (for examples, see  in \cite{VNC98,VE02,VK08}).

In this paper, we  thoroughly study gravitational lensing due to Schwarzschild black holes, which exterior 
gravitational field is described by the line element
\begin{eqnarray}
ds^2&=&\lt(1-\frac{2M}{r}\rt)dt^2- \lt(1-\frac{2M}{r}\rt)^{-1} dr^2 \nonumber \\
      &-&r^2\lt(d\vartheta^2+\sin^2 \vartheta d\phi^2\rt) \text{,}
   \label{SchSpaceTime}
\end{eqnarray}
where the real  constant parameter $M$ is the  ADM mass. The radii of event horizon
and photon sphere of a Schwarzschild black hole are  given by  $R_{eh} = 2 M$ and $R_{ps} = 3 M$, respectively.  $R_{eh}$ is also
called the Schwarzschild radius.

The bending  angle $\hat{\alpha}$ for a light ray with the  closest distance of approach $r_o$ is given by
 \cite{Wei72})
\begin{equation}
    \hat{\alpha}\lt(r_o\rt) = 2 \  {\int_{r_o}}^{\infty}
                         \frac{dr}{r \  \sqrt{\lt(\frac{r}{r_o}\rt)^2  \lt(1-\frac{2M}{r_o}\rt)
                        -\lt(1-\frac{2M}{r}\rt)}   } - \pi 
   \label{AlphaHatR0}
\end{equation}
and  the impact parameter $J$ of the light ray is expressed by
\begin{equation}
J\lt(r_o\rt) = r_o \lt(1-\frac{2M}{r_o}\rt)^{-1/2} .
       \label{ImpParaR0}
\end{equation}
Defining  a  dimensionless radial distance $\rho$ in terms of the Schwarzschild radius $2M$ by equation
\begin{equation}
  \rho = \frac{r}{2M} 
   \label{rho}
\end{equation}
(for $r=r_o, \rho=\rho_o$),  we \cite{VE00,VNC98} expressed the deflection angle $\hat{\alpha}\lt(\rho\rt)$ and the impact parameter
 $J\lt(\rho\rt)$, respectively, by 
\begin{equation}
  \hat{\alpha}\lt(\rho_o\rt) = 2 \  {\int_{\rho_o}}^{\infty}
 \frac{d\rho}{\rho  \  \sqrt{\lt(\frac{\rho}{\rho_o}\rt)^2 
 \lt(1-\frac{1}{\rho_o}\rt)
-\lt(1-\frac{1}{\rho}\rt)}} - \pi 
   \label{AlphaHatRho0}
\end{equation}

and
\begin{equation}
  J\lt(\rho_o\rt) = 2M  \rho_o \lt(1-\frac{1}{\rho_o}\rt)^{-1/2}.
     \label{ImpParaRho0}
\end{equation}

For computations of  magnifications of images, the  first derivative
of  deflection angle $\hat{\alpha}$ with respect to $\theta$ is needed, which is
given by \cite{VNC98,VE00}
\begin{equation}
\frac{d\hat{\alpha}}{d\theta} =  \hat{\alpha}'\lt(\rho_o\rt)
                               \frac{d\rho_o}{d\theta} ,
    \label{DAlphaByDTheta}
\end{equation}
where the first and second factors on right side of this equation are, respectively, given by
\begin{widetext}
\begin{equation}
\hat{\alpha}'\lt(\rho_o\rt) = \frac{3-2\rho_o}{{\rho_o}^2\lt(1-\frac{1}{\rho_o}\rt)}
{\int_{\rho_o}}^{\infty}
 \frac{\lt(4 \rho - 3\rt) d\rho}
{\lt(3 - 2 \rho\rt)^2 \  \rho \   \sqrt{\lt(\frac{\rho}{\rho_o}\rt)^2 
 \lt(1-\frac{1}{\rho_o}\rt)
-\lt(1-\frac{1}{\rho}\rt)}} 
 \label{DAlphaHatByRho0}
\end{equation}
\end{widetext}
and
\begin{equation}
 \frac{d\rho_o}{d\theta} = 
     \frac{  \rho_o \lt(1-\frac{1}{\rho_o}\rt)^{3/2}
        \sqrt{1-\lt(\frac{2M}{D_{d}}\rt)^2 {\rho_o}^2 \lt(1-\frac{1}{\rho_o}\rt)^{-1}}}
        {\frac{M}{D_{d}} \lt(2\rho_o-3\rt)} \text{.}
      \label{DX0ByDTheta}
\end{equation}

Time delays  for images of   gravitational lensing are 
given by 3 terms: the first and second terms with positive sign are, respectively, the travel time of the light 
from the source to the point of closest approach and from that point  to the observer, and the third term 
with a minus sign is  the  light travel time from the source to the observer in
the absence of any gravitational field. 
Solving null geodesic equations for general static spherically symmetric
spacetime, Weinberg  in his classic book \cite{Wei72} obtained the time required for light to travel from a source at 
coordinates $\{r, \vartheta, \pi/2, \varphi=\varphi_1\}$ to the closest point of approach (to the lens) at coordinates 
$\{r_0, \vartheta, \pi/2, \varphi=\varphi_2\}$. Using this result, time delay of images of Schwarzschild lensing
can be   expressed as (see  Eqs. (23)-(25) in \cite{VK08})
\begin{equation}
  \tau\lt(\rho_0\rt) 
      = 2M 
         \lt[         
             {\int_{\rho_0}}^{{\cal X}_s} \frac{d\rho}{f\lt(\rho\rt)} + {\int_{\rho_0}}^{{\cal X}_o} \frac{d\rho}{f\lt(\rho\rt)}          
          \rt]  
       - D_s \sec\beta                                        
    \label{tau}
\end{equation}

with
\begin{equation}
    {\cal X}_s =  \frac{D_s}{2M} \sqrt{   \lt(\frac{D_{ds}}{D_s}\rt)^2+\tan^2\beta  } \text{,} ~~~~~~
    {\cal X}_o =  \frac{D_d}{2M} \text{,} 
\end{equation}
and

\begin{equation}
    f\lt(\rho\rt)={ \sqrt{ \lt(1-\frac{1}{\rho}\rt)^{2}-\lt(\frac{\rho_0}{\rho}\rt)^2\lt(1-\frac{1}{\rho}\rt)^{3}\lt(1-\frac{1}{\rho_0}\rt)^{-1}}}  \text{.}                             
    \label{f}
\end{equation}
Time delay of a gravitationally lensed image may be  in general positive, zero, or negative; for examples, see in our paper \cite{VK08}.
However, time delays are always positive for images of Schwarzschild lensing.

It is worth mentioning that Eq. (4.67) in a classic book on GL by Schneider {\it et al.} \cite{book1} gives time delays of gravitationally lensed images. That equation
contains an additive constant term. The authors clarified that the constant term is the same for all rays from the source plane to the observer. 
Therefore, this term cancels for computations of differential time delay between $2$ images. However, as the value for the
constant term is not obtained, that equation cannot be used to compute time delays of images. This is why we  follow the approach 
given in Weinberg's book \cite{Wei72} and  we discussed that  in this section.

\section{\label{sec:EffDefAngles} Effective deflection angles of relativistic images}

It is  important to  first  discuss in brief  a few new terms we defined in our previous paper \cite{VE00}.
Then, we will show that these definitions with some arguments, reveal  geometrical beauty
of strong field Schwarzschild lensing. This also  helps predict some results without computations and
thus  provides consistency check for   results obtained through numerical computations.

If a lens is very compact, then a light ray passing close to it will  suffer  a large deflection and therefore
will loop around the lens once, twice, thrice, or many  times (depending on the closest distance of approach
from the center of the lens) before  reaching the observer. We \cite{VE00} defined {\em relativistic images} of GL  as 
those images  which occur due to light deflections by angles $\hat{\alpha}>3\pi/2$. Similarly, for  the angular source
position $\beta=0$, we defined
{\em relativistic Einstein rings}  as those  ringed-shaped images which can  form due to light deflections by angles
$\hat{\alpha}>2\pi$. Relativistic Einstein rings are thus relativistic images for the case of $\beta=0$.
It is useful to define {\em order} of relativistic images on  each side of the optical axis.
We assign the order 1 to the outermost relativistic images on both sides of the optical axis, 2 for adjacent  inner ones,
and so on. Thus, according to this definition,   the outermost relativistic Einstein ring also has 
 the order $1$.


\begin{figure*}[tbh]
\centerline{ \epsfxsize 18cm
   \epsfbox{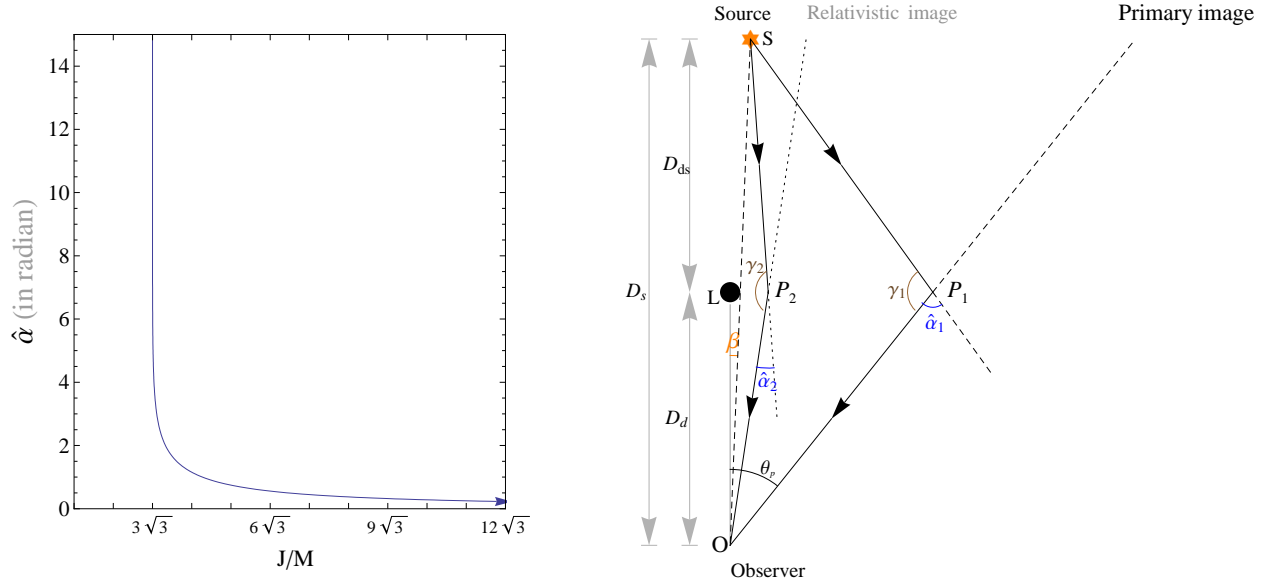}}
 \caption[ ]
{(color online). 
{\em Left}: The deflection angle $\hat{\alpha}$ is plotted against the dimensionless scaled impact parameter $J/M$.
   The arrow attached to the curve indicates that $\hat{\alpha}$  decreases  with increase in the value of 
   $J/M$.  
{\em Right}: $S$, $L$, and $O$ represent positions of the source, the lens,
  and the observer, respectively. $SP_1$ and $SP_2$ are tangents on  2 null geodesics at the
  source position, whereas $P_1O$ and $P_2O$ are  tangents on, respectively, the same pair of null geodesics at 
  the observer position. $\hat{\alpha_1}$ and $\hat{\alpha_2}$ are light bending angles, whereas $\gamma_1$ and  $\gamma_2$
  are their respective supplementary angles. $\beta$ and $\theta_p$ stand, respectively, for the angular positions of the source
  and the primary image. Angles in this schematic diagram  are greatly exaggerated. $\gamma_2 > \gamma_1 \implies 
  \hat{\alpha}_2<\hat{\alpha}_1$.
According to  the $\hat{\alpha}$ vs $J/M$ plot (on left side), $\hat{\alpha}_2<\hat{\alpha}_1$ is possible only if $SP_2$ and $P_2O$
are, respectively, tangents at the source and the observer positions on a null geodesic which loops around the lens at least
once giving rise to a relativistic image.}

\label{fig1}
\end{figure*}


\begin{figure*}[tbh]
\centerline{ \epsfxsize 20cm
   \epsfbox{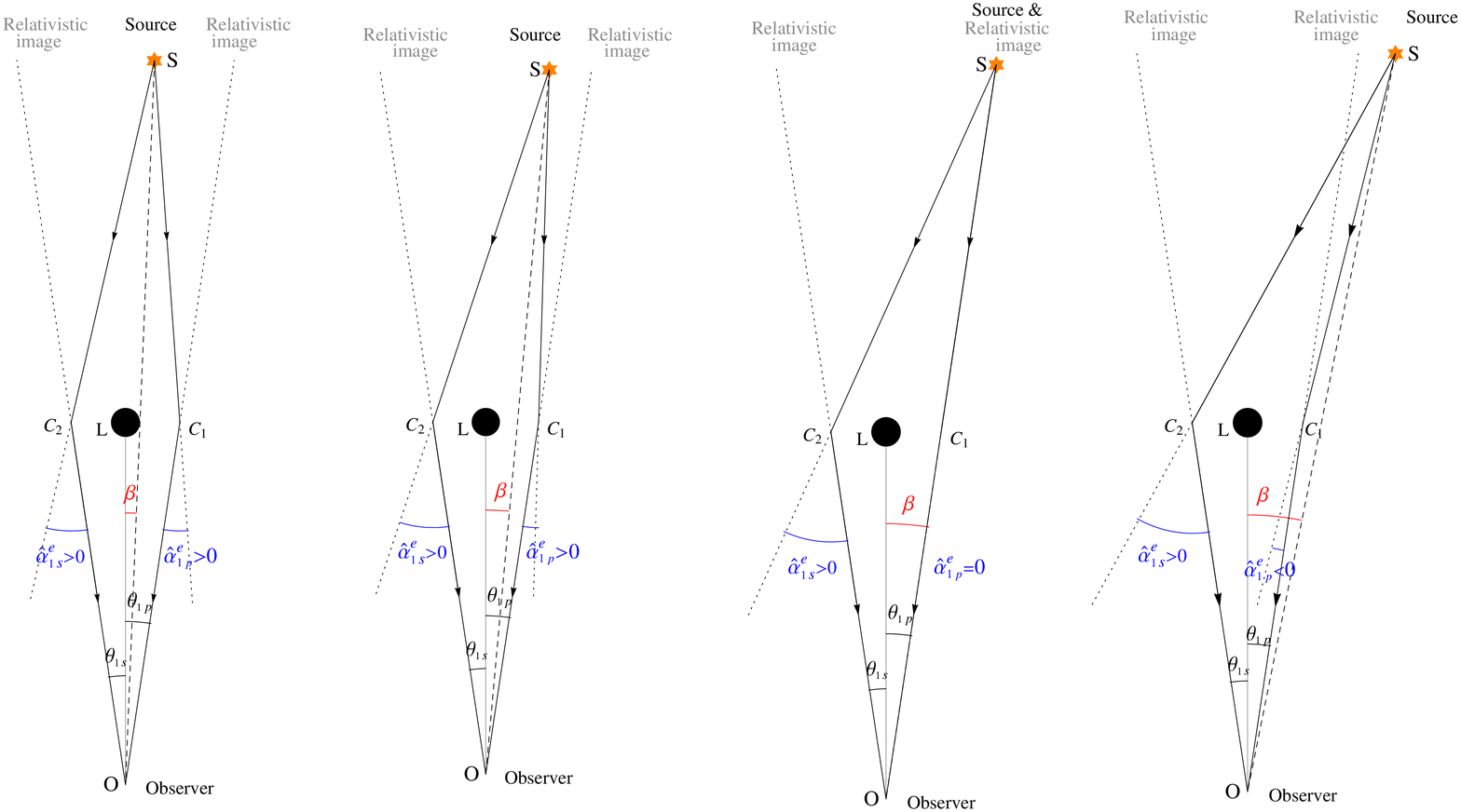}}
 \caption[ ]
  {(color online). Schematic diagram showing the variation in  effective deflection angles of relativistic images
with respect to increase in the value of angular source position $\beta$. $S$, $L$, and $O$ stand for the position
of the source, the lens, and the observer, respectively. $SC_1$ and $SC_2$ are tangents on 2 null geodesics at
$S$, and $C_1O$ and $C_2O$ are tangents on corresponding null geodesics at $O$. $\hat{\alpha}^e_{1p}$ and
$\hat{\alpha}^e_{1s}$ stand for effective deflection angles of relativistic images of order 1, respectively,
on  the primary and secondary image sides; $\theta_{1p}$  and $\theta_{1s}$ are their respective angular image positions.
Angles are   greatly exaggerated.

  }
\label{fig2}
\end{figure*}


In a recent paper \cite{VK08}, we  discussed that the existence of a photon sphere enclosing  a lens is a sufficient (not 
necessary) condition  for the formation of relativistic images. A  sufficiently compact lens can give rise to relativistic 
images even if the lens is not covered inside a photon sphere. Therefore, the lens 
need not  be a black hole to produce these images.

In \cite{VE00}, we defined a term {\em effective deflection angle} of a relativistic image, which we now express as
follows:
\begin{equation}
   \hat{\alpha}^{e}\lt(\rho_0\rt) = \hat{\alpha}\lt(\rho_0\rt)-2n\pi \text{,}
\label{EffDefAngle}
\end{equation}
where  $n$ is a positive integer that represents the number of loops (turns) a light ray makes  around the lens
before reaching
the observer,  and  the superscript $e$ on $\hat{\alpha}$ stands for the word {\em effective}. 
(In fact, the above equation can be also applied to primary and secondary images, because $n=0$ correspond to those images.
However, we prefer to call those  as known in the literature instead  of calling them images of 0-order.)
$\hat{\alpha}\lt(\rho_0\rt)$,
as given by Eq.  $(\ref{AlphaHatRho0})$, is the  usual (total) deflection angle for the light ray
with the closest  scaled distance of approach $\rho_0$ making $n$ loops around the lens. Thus, similar to the cases of primary and secondary
images, the effective deflection angle of a relativistic image turns out to be the angle between the tangents on the
null geodesic at the source and at the observer positions. In the following,  we now introduce new symbols we use for physical quantities
associated with primary, secondary, and relativistic images.

{\em New symbols}.--- We use subscripts $p$ and $s$, respectively, for primary and secondary images. Similarly,
subscripts $np$ and $ns$ ($n=1,2,3,...$) stand, respectively, for the images of order $n$ on the same side as the primary and 
secondary images. For example, $\hat{\alpha}^{e}_{1p}$ and $\hat{\alpha}^{e}_{1s}$, respectively, stand for effective
deflection angles of relativistic images of order $1$ on the same side as the primary and secondary images. The same applies
to symbols for angular image positions, magnifications, and time delays.

It is well-known that, on the same side as the source from the optical axis, the Schwarzschild black 
hole lensing gives rise to the  primary image which is formed  due to light deflection in weak gravitational field
without looping of the ray of light around  the lens.  On the other hand, relativistic
images on the same side as the source are produced due to  looping of light rays around the lens, which is caused by
large deflection angles $\hat{\alpha} > 3 \pi/2$
in strong gravitational field. It is natural to ask if, excluding the primary
image, there is any other image (on the same side as the source) which can form without looping of the light ray around the black hole.
Computations give no such solutions to the lens equation. Therefore, on the same side as
the source, there is only one (i.e., the primary) image which forms due to light deflection in weak field without looping of the
light ray around the lens and  there are relativistic images which arise due to looping of light rays around the lens in strong gravitational
field. In the following, we will show that a simple geometrical argument beautifully supports
these numerical results.

See Fig. 1. Assume that, on the same side as the source from the optical axis,  two light rays emitted from the source $S$ reach 
the observer $O$ without  looping  around the lens. $SP_1$ and $SP_2$ are tangents on 2 null geodesics
at the source position, and, similarly, $P_1O$ and $P_2O$ are, respectively, tangents on those null geodesics at the
observer position. $\hat{\alpha}_1$ and $\hat{\alpha}_2$ are deflections angles
corresponding to 2 light rays we considered and $\gamma_1$ and $\gamma_2$ are their respective supplementary 
angles. The schematic diagram (right  of Fig. 1) shows that  $\gamma_2 > \gamma_1$. This implies that
$\hat{\alpha}_2 < \hat{\alpha}_1$, which is not allowed according to the $\hat{\alpha}$ vs $J/M$ graph (see the left  of Fig. 1);   
i.e., a decrease in the impact parameter should increase $\hat{\alpha}$. Thus if
the light   path $SP_1O$ is allowed, then
$SP_2O$ is not allowed and therefore we conclude that there can be only the primary image on the same side as
the source without the light ray looping around the lens. A similar argument also demonstrates that there can be
only one image (i.e., secondary) on the opposite side  from the source without a light ray going around  the lens.

Now consider that  $SP_2$ and $P_2O$ represent tangents, respectively, at the source and
the observer position on a null geodesic that loops around the lens once before reaching the
observer. Therefore, $\hat{\alpha}_2 = \hat{\alpha}^e_{1p} < \hat{\alpha}_1$, which also reflects in numerical
computations in next section. For a given value of $\beta$, the schematic diagram also shows that the effective deflection angle of relativistic
images on the same side as the primary image decrease with increase in its order. Similarly, for a given angular source position, the effective
deflection angles of relativistic images (on the same side as the secondary) decreases with the increase in its order.
These conclusions based on simple geometrical analysis are reflected in  results of our  numerical computations
in the next section.

In Fig. 2, we show  that effective deflection angles of relativistic images on the same
side as the primary image can be positive, zero, or negative depending on the value of the angular source position.
However, for relativistic images on the same side as the secondary image, effective deflection angles are always
positive. Consider the first order relativistic images on both side of the optical axis; i.e., one on the primary
image side and the other on the secondary image side. $SC_1$ and  $C_1O$ are, respectively, tangents on null geodesics
(giving rise to the $1^{st}$ order relativistic image on the primary image side) at the source and observer positions; 
$C_1$ is their point of intersection. Similarly, $SC_2$ and $C_2O$ are, respectively, tangents on null geodesics
(giving rise to the $1^{st}$ order relativistic image on the secondary  image side) at the source and observer positions; 
$C_2$ is their point of intersection. For small angular source position $\beta$,  the effective deflection angle 
$\hat{\alpha}^e_{1p}>0$ (see the extreme left diagram). As the value of $\beta$ increases, the value of $\hat{\alpha}^e_{1p}$ decreases
to zero value (see the second and third diagrams from the left). When $\hat{\alpha}^e_{1p}=0$, the angular 
source position of this relativistic image and the source coincide. We denote this critical value of the
angular source position as $\beta_{1c}$. (subscript $c$ stands for the word {\em critical} and $1$ stands for the $1^{st}$ order relativistic image.) 
A further increase in the value of $\beta$ makes the effective deflection angle $\hat{\alpha}^e_{1p}<0$
and its value  keep decreasing with  increase in the value of $\beta$. On the other hand, the effective deflection angles
of relativistic images on the same side as the secondary  image  increases with the increase in the value of $\beta$.
These important conclusions based on simple geometrical analysis  also appear in our numerical  results  in the next section.
Therefore, the geometrical analysis using Fig. 1 and Fig. 2  also supports correctness of our computations.



\begin{figure*}[tbh]
\centerline{ \epsfxsize 16cm
   \epsfbox{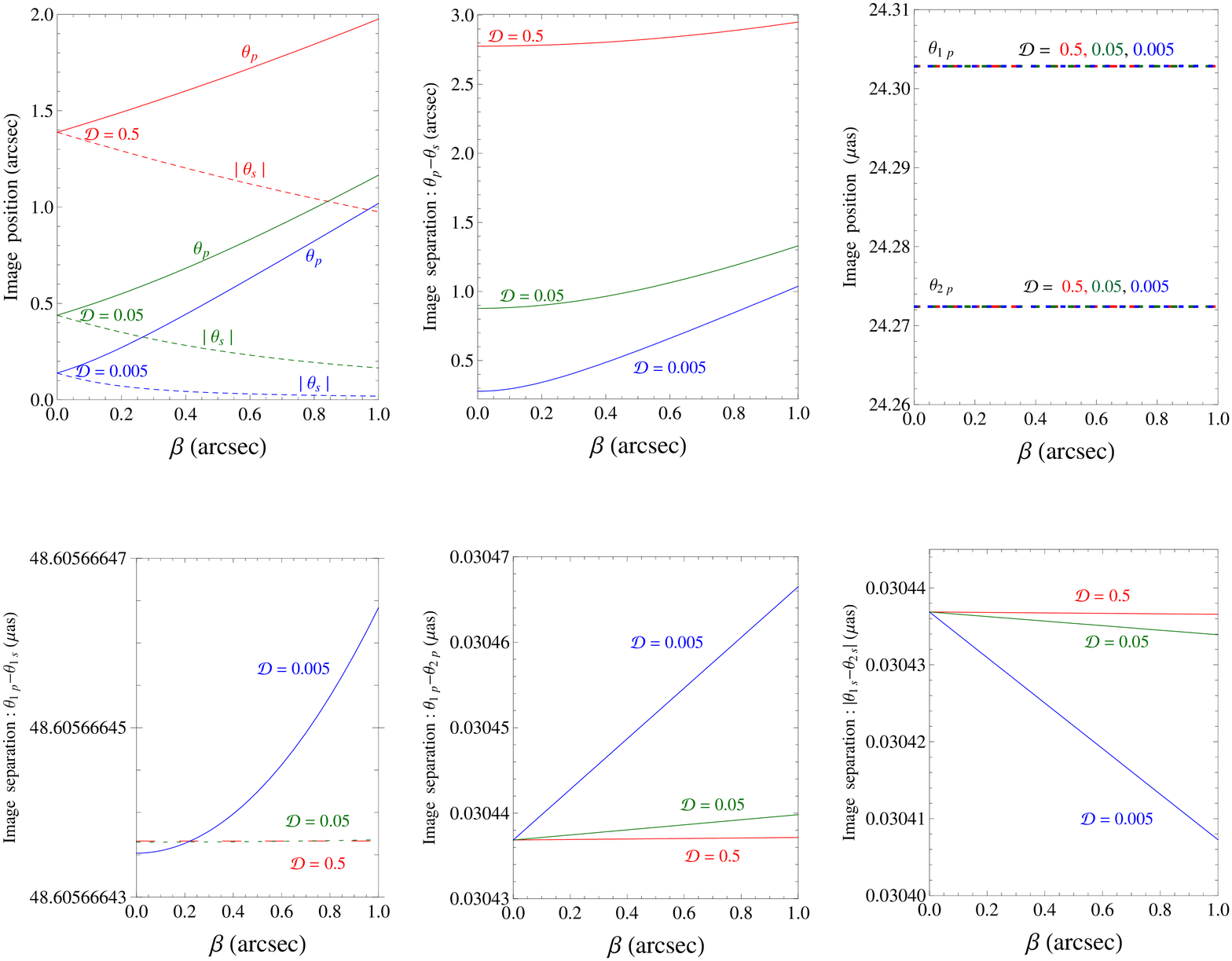}}
 \caption[ ]{(color online).
{\em Top  left and middle}:  The  angular  positions of primary images $\theta_p$, secondary images $|\theta_s|$, and their separations
$\theta_p-\theta_s$ are plotted
 against the angular source position $\beta$ for ${\cal D}=0.5,0.05$ and $0.005$. 
{\em Top right}: The angular
 positions of  relativistic images (on the same side  as the primary image) of the first order $\theta_{1p}$ and
of the second order $\theta_{2p}$ are plotted against $\beta$ for the same values of ${\cal D}$ as in the 
figures on left.  The curves for  $\theta_{1p}$ (for different values of  ${\cal D}$)
intersect for $\beta_{1c} \approx 24.3028$  $\mu as$, whereas those for $\theta_{2p}$  intersect for  $\beta_{2c} \approx 24.2724$
$\mu as$. 
{\em Below}:  The angular separations among relativistic images vs the angular source position $\beta$ are plotted for 
${\cal D}=0.5,0.05$ and $0.005$. $\theta_{np}$ and $\theta_{ns}$ ($n=1,2$) stand for angular positions of relativistic images
on the primary and the secondary image sides, respectively.
The Galactic MDO is modeled as the 
Schwarzschild lens, which has  mass $M= 3.61 \times 10^6 M_{\odot}$
and is situated at  the distance $D_d =  7.62$ {\em kpc}  so that  $M/D_d \approx 2.26 \times  10^{-11}$.
 
}
\label{fig3}
\end{figure*}




\section{\label{sec:GalMDO} Gravitational lensing by the Galactic MDO}


\begin{figure*}[tbh]
\centerline{ \epsfxsize 16cm
   \epsfbox{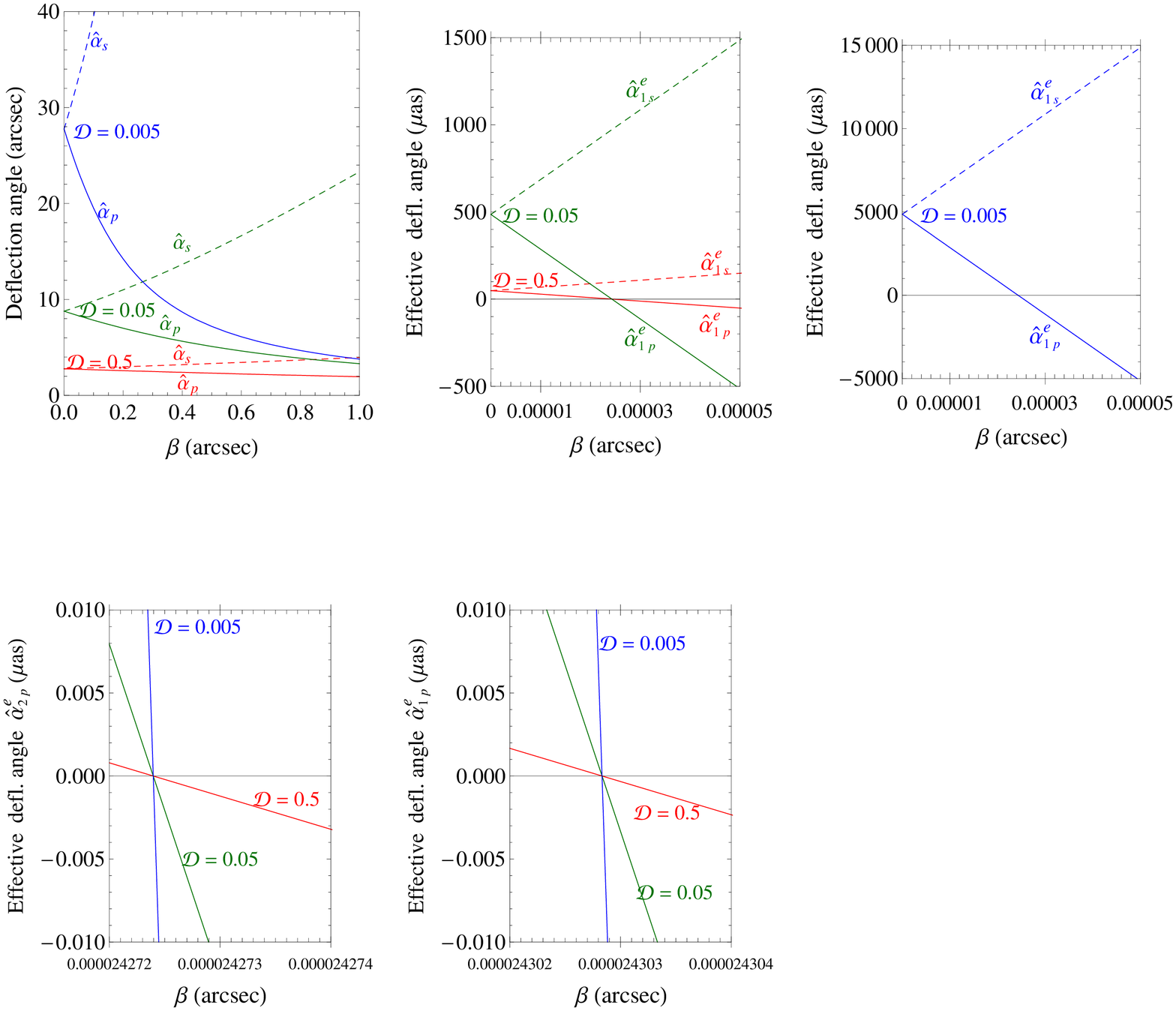}}
 \caption[ ] {(color online).
 {\em Top left}: The  deflection angles of primary images $\hat{\alpha}_p$ and secondary
 images  $\hat{\alpha}_s$ are plotted against the angular source position $\beta$ for ${\cal D}=0.5,0.05$ 
 and $0.005$.
 {\em Top middle and right}: The effective deflection angles of the first order relativistic images on the same side as
 the primary image $\hat{\alpha}^e_{1p}$ and on the same side as the secondary  image $\hat{\alpha}^e_{1s}$
 are plotted against $\beta$ for ${\cal D}=0.5, 0.05$ and $0.005$.
{\em Below left and right}: The effective deflection angles of the second order $\hat{\alpha}^e_{2p}$ (left)
and of the first order $\hat{\alpha}^e_{1p}$ (right), both on the same side as the primary image, are
plotted against $\beta$ for ${\cal D}=0.5,0.05$ and $0.005$ in the vicinity of zero effective deflection angle.
$\beta_{2c} \approx 24.2724 \ \mu as$ and $\beta_{1c} \approx 24.3028 \ \mu as$, where  $\beta_{1c}$ and $\beta_{1c}$ are, respectively, critical angular source
positions for the second and first order relativistic images. The gravitational lens is the same as for the Fig.~3.
}
\label{fig4}
\end{figure*}


\begin{figure*}[tbh]
\centerline{ \epsfxsize 16cm
   \epsfbox{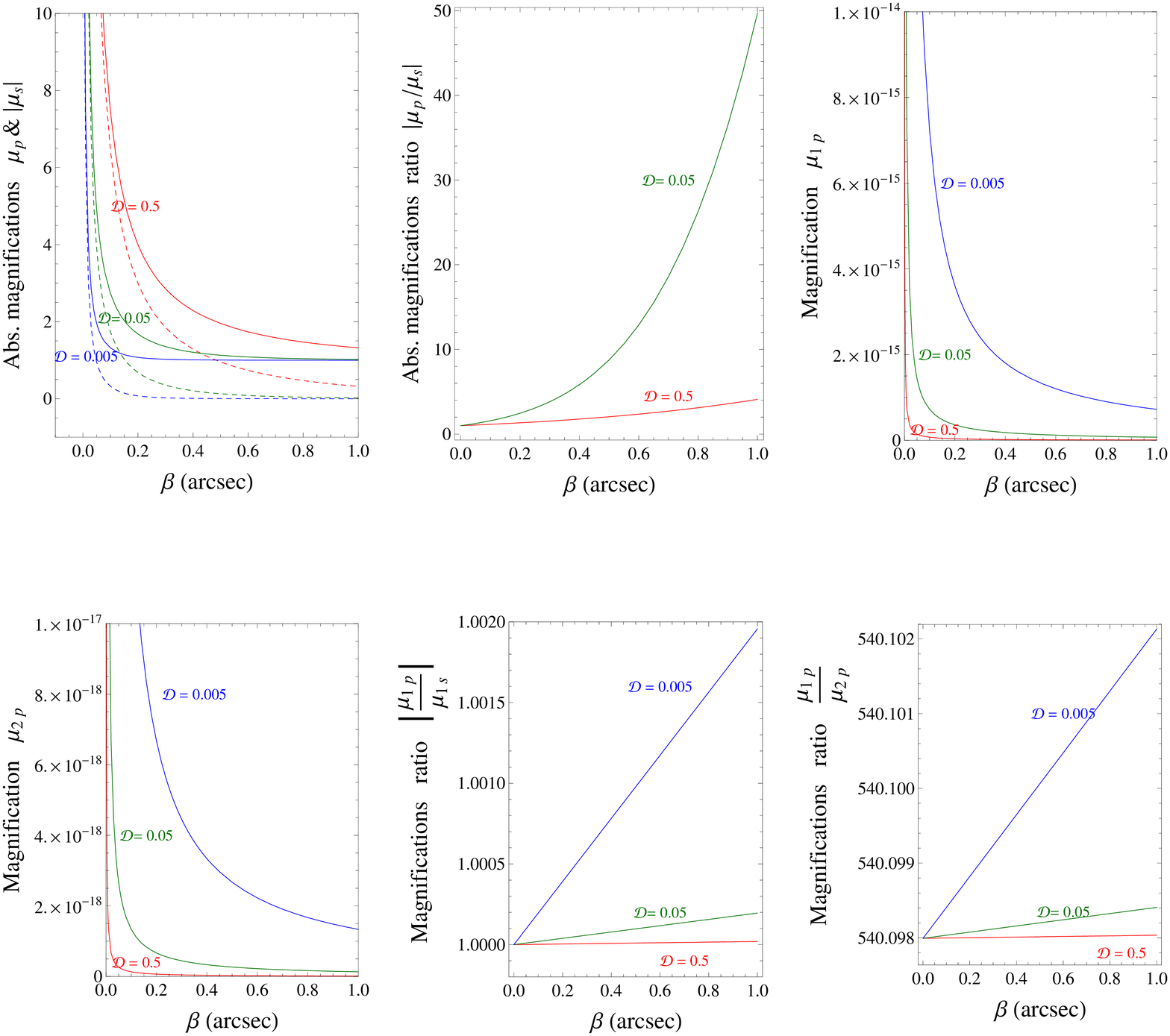}}
\caption[ ] {(color online). 
{\em Top left and middle}:  The magnifications of primary images $\mu_p$, the absolute magnifications of secondary images $|\mu_s|$, and 
their ratios $|\mu_p/\mu_s|$ are plotted against the angular source position $\beta$ for different values of  ${\cal D}$.
{\em Top right and below left }: The magnifications of relativistic images (on the same side as the primary image) of the first order $\mu_{1p}$ and
the second order $\mu_{2p}$ are plotted against the angular source position $\beta$ 
for ${\cal D}=0.5,0.05$ and $0.005$. 
{\em Below middle and right}: The magnifications ratios $|\mu_{1p}/\mu_{1s}|$ (where $\mu_{1s}$ stands for magnification of the first order relativistic
image on the same side as the secondary image) and $\mu_{1p}/\mu_{2p}$ vs $\beta$ are plotted for the same values of ${\cal D}$ as in the figure on below 
left. The lens is the same as for the Fig. 3.
}
\label{fig5}
\end{figure*}



\begin{figure*}[tbh]
\centerline{ \epsfxsize 16cm
   \epsfbox{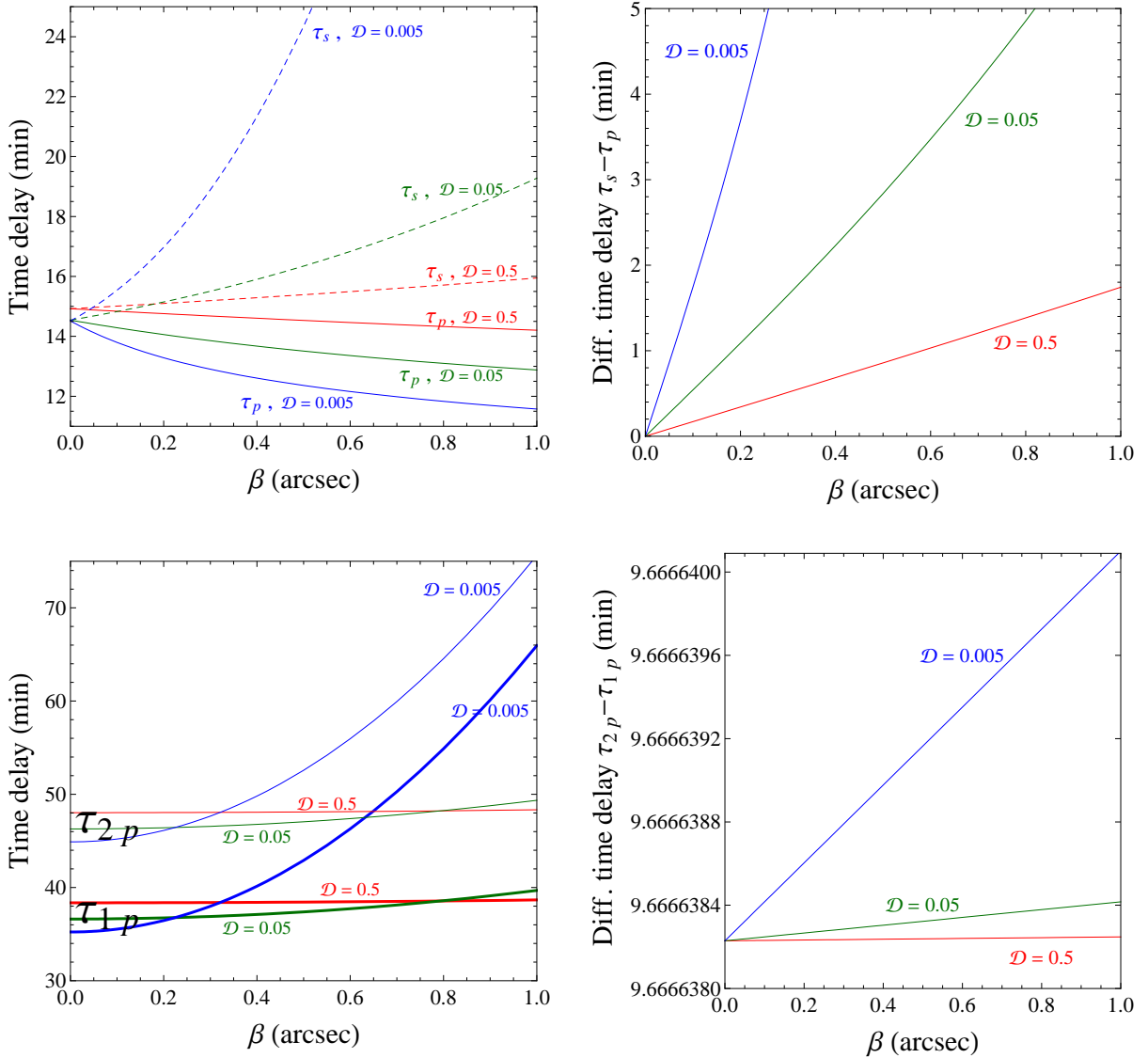}}
\caption[ ] {(color online).
 {\em Top}:  The time delays of primary images $\tau_p$ and secondary images $\tau_s$, and the differential time delays of the
  secondary images with respect to their respective primary images (i.e., $\tau_s - \tau_p$) are plotted against the angular source position $\beta$ for
 ${\cal D}=0.5,0.05$ and $0.005$.
{\em Below}: The time delays of relativistic images (on the same side as the primary image) of the first order $\tau_{1p}$,
  second order $\tau_{2p}$, and the differential time delays (i.e., $\tau_{2p} - \tau_{1p}$) are plotted against  $\beta$ for the same values  ${\cal D}$ as in figures on top.
The lens is the same as for the Figs. 3 through 5.
}
\label{fig6}
\end{figure*}



\begin{figure*}[tbh]
\centerline{ \epsfxsize 14cm
   \epsfbox{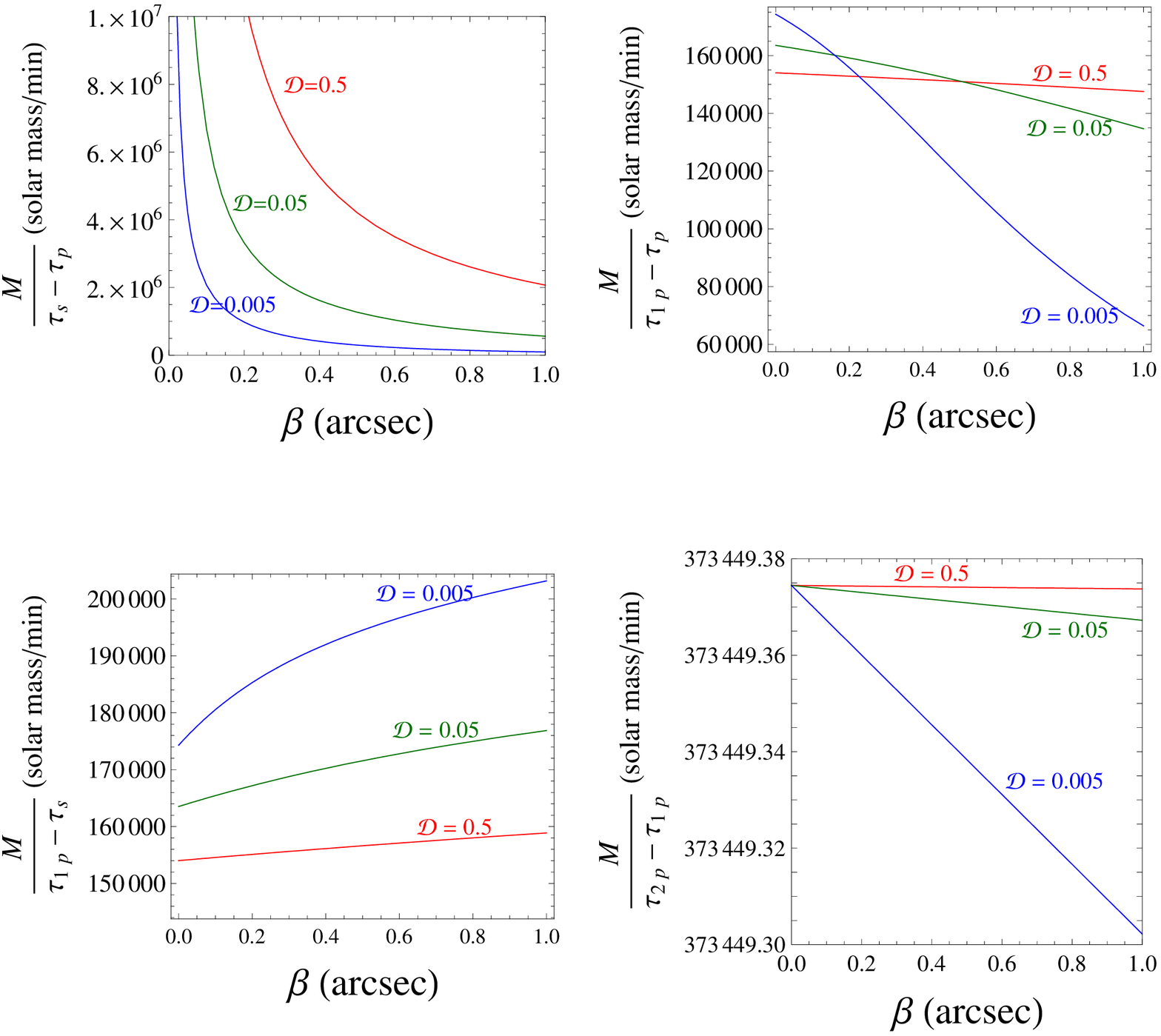}}
 \caption[ ] { (color online).
The  ratios  of mass $M$ of the lens  to   differential time delays of  images
are plotted against the angular source position $\beta$ for ${\cal D}=0.5,0.05$ and $0.005$.
$\tau_p$ and $\tau_s$ stand for time delays of primary and secondary images, respectively. 
$\tau_{1p}$ and $\tau_{2p}$, respectively, represent time delays of relativistic images (on the
same side as the primary image) of orders $1$ and $2$. The gravitational lens is the same as for
the Figs. 3 through 6.

  } 
\label{fig7}
\end{figure*}


In this section, we model the MDO at the center of our Galaxy (the Milky Way) as a Schwarzschild black hole lens and study 
point source GL in a great detail. The MDO has the mass $M= 3.61 \times 10^6 M_{\odot}$ and is at the distance
$D_d =  7.62$ {\em kpc} from us \cite{Eisenetal05}. Therefore,  $M/D_d \approx 2.26 \times  10^{-11}$ (note that $M \equiv M G/c^2$). 
In a recent paper \cite{VK08},  we already obtained angular positions,  deflection angles, magnifications, and
time delays for primary and secondary images for several values of angular source position $\beta$ for 
${\cal D}\equiv D_{ds}/{D_s}=0.5$ (i.e., when the lens is situated halfway between the observer and the source).
We also computed differential time delays of secondary images with respect to primary images. For comparison and 
continuity in discussion, we use  those results in this paper and also put  those in Table I in the Appendix of
this paper.
Now considering ${\cal D}=0.5$ and using {\em Mathematica}, we numerically solve the  gravitational lens equation 
$(\ref{GravLensEqn})$ for a large number of values for $\beta$ and obtain image positions for first and second order
relativistic images on both sides of the optical axis. We further obtain  deflection angles, magnifications,
time delays, and differential time delays for these images. Using Eq. $(\ref{EffDefAngle})$, we obtain effective deflection
angles for these relativistic images. We put these results in Tables II and III. Though, we computed for a large number of values
for $\beta$, we put only a few data in tables; however, we use all those for figures. We further  repeat the entire computations
for primary and secondary as well as relativistic images for  ${\cal D}=0.05$ and $0.005$ to see the effects of changes in 
image positions, deflection/effective deflection angles, magnifications, time delays, and differential time delays due
to change in the lens-source distance. (With the observer-lens distance $D_d$ fixed, a decrease in the value of ${\cal D}$ decreases
the source-lens distance.) Throughout our computations in this paper, we never take either weak or strong  gravitational field 
approximation and therefore our computations and hence results are {\em exact} in this sense. In the following paragraphs, we will
now discuss  results for  GL by the Galactic MDO.

In Fig. 3, we plot the (absolute) angular positions (measured from the optical axis) of primary and secondary images, and their separations
against the angular source 
position $\beta$ for ${\cal D} = 0.5, 0.05$ and $0.005$. As it is well-known that, for a given value of ${\cal D}$, 
the angular positions of primary and secondary images, respectively, increase and decrease with an increase in the value
of $\beta$. For a given value of $\beta$, the angular positions of primary as well as secondary images
increase with an increase in the value of ${\cal D}$. The angular radius of Einstein ring increases with increase in the value
of ${\cal D}$. The angular separation between primary and secondary images increases with increase in $\beta$ and ${\cal D}$.
The angular source positions of relativistic images are very insensitive to changes in 
the values of $\beta$ and ${\cal D}$. The angular position of the first order relativistic images on both sides of the optical axis
have extremely close values; however, $\theta_{1p} > |\theta_{1s}|$ for all values of $\beta$, excluding, of course, at $\beta=0$ for which 
$\theta_{1p} = |\theta_{1s}|$. The same is true for any pair of second or higher order relativistic images. As $\theta_{np}$ and
$|\theta_{ns}|$ (for the same value of $n$) have extremely close values, we  plot image  positions of relativistic images  only on the 
same side as the primary image.  For theoretical interest, it is worth investigating variation in the value of 
$\theta_{np}$ and $|\theta_{ns}|$ with changes in the value of $\beta$ and ${\cal D}$, though variations are extremely small.
For relativistic images of a given order $n$ and for a given value of ${\cal D}$, the values in $\theta_{np}$ and $|\theta_{ns}|$, 
respectively, increase and decrease
with the increase in the value of $\beta$, though the increase and decrease are extremely small. However,  their dependence on ${\cal D}$
is much more fascinating. For $\beta=0$ or a small
value, $\theta_{1p}$ is greater for a greater value of ${\cal D}$. As $\beta$ increases, there is situation when $\theta_{1p}$ is the 
same for all values of ${\cal D}$. At this critical angular source position $\beta_{1c} \approx 24.3028$ {\em microarcsec} ($\mu as$),
the effective deflection angle $\hat{\alpha}^e_{1p}=0$ for all values of ${\cal D}$. For a further increase in the value
of $\beta$, $\theta_{1p}$ is smaller for a greater value of ${\cal D}$. The same happens for any relativistic images on the same 
side as the primary image; however, critical angular source position $\beta_{nc}$ decreases with an increase in  the order $n$ of the image; for example,
$\beta_{2c} \approx 24.2724$ $\mu as$.  On the other hand, for any fixed value of $\beta$, image positions of relativistic images (on the secondary image side) 
$|\theta_{ns}|$  always increase with an increase in ${\cal D}$. 
The angular separation between relativistic images of the first order (i.e, $\theta_{1p} -\theta_{1s}$) increases 
with an increase in the value of $\beta$ (for a fixed ${\cal D}$). The increase rate of the angular separation with increase in $\beta$ is higher for lower value of
${\cal D}$. For $\beta = 0$ or a small value, this angular separation is higher for a higher value of ${\cal D}$. (This is 
qualitatively similar to the case of image separation between primary and secondary images.) However, for large value of
$\beta$, the angular separation $\theta_{1p} -\theta_{1s}$ is higher for lower value of ${\cal D}$.
For any two values of  ${\cal D}$, there exist a $\beta$ at which  values of $\theta_{1p} -\theta_{1s}$ are equal for both
${\cal D}$. The angular separations between outermost $2$ relativistic images (both of them either on the primary or on the 
secondary image side) have just opposite qualitative dependence on $\beta$ and ${\cal D}$ in the following sense. 
$\theta_{1p} -\theta_{2p}$ increases with increase in $\beta$ (for a fixed ${\cal D}$), but decreases with increase in
${\cal D}$ (for a fixed $\beta$). On the other hand,  $|\theta_{1s} -\theta_{2s}|$ decreases with increase in $\beta$
(for a fixed ${\cal D}$), but increases  with increase in ${\cal D}$ (for a fixed $\beta$). The increase/decrease rate 
with change in the value of $\beta$ (for a fixed ${\cal D}$) is smaller for higher value of ${\cal D}$.
The variations in angular separations between relativistic images are very small with respect to changes in $\beta$ and 
${\cal D}$. Among angular separations $\theta_{1p} -\theta_{1s}$, $\theta_{1p} -\theta_{2p}$, and $|\theta_{1s} -\theta_{2s}|$,
the first one is the least sensitive to those changes. As relativistic images would be observationally important for very small values of
$\beta$, we conclude that angular separations between relativistic images of our interest are extremely insensitive to change
in the value of ${\cal D}$.

In Fig.~4, we show  variations in deflection angles for primary and secondary as well as effective deflection angles for 
relativistic images with respect to changes in values for $\beta$ and ${\cal D}$. It is known that for a given value
of ${\cal D}$, the deflection angles for primary and secondary images, respectively, decrease and increase with an increase 
in the value of $\beta$. For a fixed value of $\beta$, deflection angles for  these images increase with decrease 
in the value of ${\cal D}$. Like primary and secondary images, the deflection angles of relativistic images are always
positive, and the same is true for   effective deflection angles of relativistic images on the secondary image side. However,
for a fixed value of $M/D_d$, the effective deflection angles of relativistic images on the same side as the primary image
decrease with an increase in the value of $\beta$, and  can be positive, zero, or negative depending on the value of $\beta$.
For a relativistic image (of any order) on the same 
side as the primary image, there exists a critical value of $\beta$ for which the deflection angle is zero. 
These results are as expected from the schematic diagram in Fig.~2. Our numerical
computations give $\beta_{2c} \approx 24.2724$ $\mu as$ and  $\beta_{1c} \approx 24.3028$ $\mu as$ showing that 
 $\beta_{2c} <  \beta_{1c}$.
The effective deflection angle is positive, zero, and negative, respectively, for the angular source position less than, equal to, and
greater than the critical value of the angular source position.  The critical source positions are independent of the value of ${\cal D}$.
As expected from the schematic diagram (see Fig.~2), our computations also show that effective deflection angles of relativistic images
on the secondary image side are always positive and
increase with increase in the value of $\beta$ (for fixed ${\cal D}$).

In Fig. 5, we show  changes in (absolute) magnifications of primary and secondary images as well as relativistic images with
changes in the values of $\beta$ and ${\cal D}$.  Images on the same side as the source and opposite side  from 
the source have, respectively, positive and negative magnifications and therefore have, respectively, positive and negative parities.
The (absolute) magnifications of primary, secondary, and relativistic images decrease with an increase in the value of
angular source position $\beta$. However, there are 2 important differences between absolute magnifications of primary-secondary 
pair and relativistic images. First, the absolute
magnifications of relativistic images are extremely small and  decrease much faster than those of primary and secondary images with an increase in the angular source
position $\beta$. Secondly, as opposed to the case of primary and secondary images, for a fixed value of $\beta$, the absolute magnifications
of relativistic images increase with decrease in the value of ${\cal D}$. Therefore, it would be easier to observe relativistic images
of those sources which are  relatively nearer to the lens.
The absolute magnifications of relativistic images of the same order on each side of the optical axis have extremely close values
(images on the same side as the primary image have though slightly higher value than images on the same side as the secondary image). This is why we 
plot only for relativistic images on the primary image side.
For a fixed value of ${\cal D}$, the ratios of absolute magnifications of primary and secondary images $|\mu_p/\mu_s|$, and of relativistic images 
$|\mu_{1p}/\mu_{1s}|$ and $\mu_{1p}/\mu_{2p}$ increase with increase in the value of $\beta$. However, for a given value $\beta$, these decrease with an 
increase in ${\cal D}$.
Compared to the (absolute) magnifications ratio of   primary and secondary images,
the ratios of (absolute) magnifications of relativistic images are much less sensitive to changes in $\beta$ and
${\cal D}$ .

In Fig. 6, we first plot time delays of primary and secondary images, and differential time delays of secondary images
with respect to their respective primary images against the angular source position $\beta$ for ${\cal D} = 0.5, 0.05$, and $0.005$.
For Einstein rings ($\beta=0$ case), time delay decreases  with a decrease in the value of ${\cal D}$. For any given
value of  ${\cal D}$, the time delays of primary and secondary images, respectively, decrease and increase with increase 
in the value of $\beta$. Similarly, for any given value of $\beta$, time delay of a primary image decreases with
a decrease in the value of  ${\cal D}$. However, there is no such simple dependence for the time delays of secondary images
on  ${\cal D}$.  For a small value of $\beta$, time delay of secondary image is smaller for smaller  ${\cal D}$ and
the difference decreases as $\beta$ increases. For a certain value of $\beta$, time delays for secondary image for 2 different
values of  ${\cal D}$ become equal. For a further increase in $\beta$, time delays for secondary images are higher for lower
value of  ${\cal D}$ and the difference keeps increasing with increase in $\beta$. For a fixed value of  ${\cal D}$,
the differential time delay of secondary image with respect to the primary image increases with increase in $\beta$.
However, for a fixed value of $\beta$, this differential time delay increases with decrease in  ${\cal D}$.
We now plot time delays of relativistic images of the first and  second orders (both on the same side as the primary
image), and  the differential time delay of the first with respect to the second against $\beta$ for  ${\cal D} = 0.5,
0.05$, and $0.005$. The time delays of relativistic images of the same order on each side of the optical axis have
extremely close values (images on the same side as the primary image though have lower values than images on the same side as
the secondary image). This is why we  plot only for relativistic images on the primary image side.
The differential time delay ($\tau_{2p}-\tau_{1p}$) has  simple  dependence on $\beta$ and  ${\cal D}$. For a fixed value of  ${\cal D}$,
the differential time delay increases with an increase in the value of  $\beta$; however, for any fixed value of $\beta$,
the differential time delay increases with a  decrease in the value of  ${\cal D}$. For fixed $\beta$ and  ${\cal D}$, time
delays of relativistic images increase with increase in the order; i.e., inner relativistic images have higher time delays
relative to outer relativistic images. For a given value of  ${\cal D}$ and order $n$, time delay of a relativistic image
increases with increase in the value of $\beta$; the rate of increase is higher for lower values of  ${\cal D}$.
However, for a fixed value of $\beta$ and order $n$ of relativistic image, the dependence of time delays on  ${\cal D}$ is
not so simple. For $\beta=0$ or a small value, the time delay of a relativistic image of a given order is smaller
for a smaller value  ${\cal D}$; however, for a certain value of $\beta$, both equal and as $\beta$ increases further,
time delay is higher for a lower value of  ${\cal D}$.

In Fig. 7, we plot ratios of the mass of the lens to differential time delays among images against the angular source 
position $\beta$ for  ${\cal D}=0.5, 0.05$, and $0.005$. We choose 4 differential time delays: (a) differential time
delay of secondary image with respect to the primary image; i.e., $\tau_s-\tau_p$, (b) differential time delay of the
first order relativistic image (on the same side as the primary image) with respect to the primary image, i.e.,
$\tau_{1p}-\tau_p$, (c) differential time delay of the
first order relativistic image (on the same side as the primary image) with respect to the  secondary image, i.e.,
$\tau_{1p}-\tau_s$, and (d) differential time delay of the second order relativistic image with respect to the first
order relativistic image (both on the same side as the primary image), i.e., $\tau_{2p}-\tau_{1p}$. We do not consider
some other combinations for differential time delays for obvious reasons; for example, we do not use $\tau_{1s}-\tau_{1p}$,
because these are too small  (see Tables 2 and 3) to be  measured possibly in several  decades to come. For a fixed
value of ${\cal D}$, ratios $M/(\tau_s-\tau_p), M/(\tau_{1p}-\tau_p)$, and the ratio $M/(\tau_{2p}-\tau_{1p})$ decrease and
$M/(\tau_{1p}-\tau_s)$ increase with increase in the value of $\beta$. For any given value of
$\beta$, $M/(\tau_s-\tau_p)$  and $M/(\tau_{2p}-\tau_{1p})$ decrease and $M/(\tau_{1p}-\tau_s)$ increases with a decrease
in the value of ${\cal D}$; however, dependence of $M/(\tau_{1p}-\tau_{p})$ on ${\cal D}$ is somewhat complex and fascinating.
For $\beta=0$ or a small value, the ratio $M/(\tau_{1p}-\tau_p)$ is higher for smaller ${\cal D}$. As $\beta$ increases, 2 curves for
 2 different values of ${\cal D}$ intersect and hence this ratio is the same for both values of ${\cal D}$.
For a further increase in $\beta$, the ratio is now higher for higher value of ${\cal D}$. Fig.~7 shows that the ratio
$M/(\tau_{2p}-\tau_{1p})$ is the most insensitive to changes in values of $\beta$ and ${\cal D}$. 
In fact, as the relativistic images can be observed only for $\beta = 0$ or a very small value, the variation in the 
ratio $M/(\tau_{2p} -\tau_{1p})$ due to change in ${\cal D}$ is extremely small.
In the next section, we will show that this ratio is in fact extremely insensitive to change in the value of $M/D_d$ as well.
Therefore, the physical quantity $M/(\tau_{2p}-\tau_{1p})$ can be  approximately considered as a constant, which can be used
to compute  very accurate values for  masses of black holes once differential time delays $\tau_{2p}-\tau_{1p}$ are measured.

We mentioned in the first section of this paper that Bozza {\em et al.} \cite{Bozetal01} analytically obtained
approximate expressions for image positions and magnifications of relativistic images. In order to calculate 
angular  positions of these images, they first obtained an expression for effective deflection angles (though they
did not call that by this name). Here, we briefly compare our results with their by giving some examples. For the
purpose of comparison,  we consider the MDO at the center of the Milky way as the lens. This lens has
$M/D_d \approx 2.26 \times  10^{-11}$. We consider the lens to be situated halfway between the source and
the observer (i.e.,  ${\cal D} = 0.5$)  and the angular source position $\beta = 1 \mu as$. Compared to our 
results, Bozza {\em et al.}  expressions give $\approx  0.5 \% $ higher values for
each of the following: $\hat{\alpha}^e_{1p}$ (effective deflection angle of the relativistic image of order $1$
on the primary image side), $\theta_{1p}$ (angular position of the relativistic image of order $1$
on the primary image side), and $\theta_{1p}-\theta_{1s}$ (angular separation between relativistic images of
the first order). Though, percentage differences appear small, these are significant for $2$  reasons. We have
shown that the angular positions of relativistic images and their separations are extremely insensitive to changes
in the angular source position as well as lens-source distance. In view of this fact, the above percentage differences
are significant. In the next section, we show that angular separation between $2$ relativistic images can be
used to obtain very accurate value for distance of a lens. Therefore,  percentage errors in Bozza
{\em et al.} results will decrease  accuracies in determination of distance of MDOs. Secondly, observation of relativistic 
images would provide a method to test the general theory of relativity against alternative  theories of 
gravity in strong  gravitational field region. Angular positions of relativistic images of the same order and the same parity in different
theories of gravity are expected to be  very close.  Therefore, very accurate theoretical results for image positions
would be  required to  compare different theories of gravity. We now compare results for magnifications of relativistic
images due to the same lens. For ${\cal D} = 0.5$ and $\beta = 1 \mu as$, Bozza {\it et al.} result yields 
$\approx 372 \% $ higher value than our  for $\mu_{1p}$ (magnification  of the first order relativistic image
on the primary image side). This very large percentage difference appears to be due to unrealistic drastic approximation
they took in their calculation. Moreover, according to their result, the absolute magnification of relativistic
images of the same order are equal; i.e. $|\mu_{np}/\mu_{ns}| = 1$, which is obviously not correct due to the asymmetry
($\beta \neq 0$). Our results show that $|\mu_{np}/\mu_{ns}| > 1$, as expected.
For qualitative similarities between Bozza {\em et al.} and our  results, see \cite{Bozetal01}.
Bozza and Mancini \cite{BozMan04} also analytically obtained approximate expressions for differential time delays among 
relativistic images. In the next section, we show  that there are again large percentage errors in their results.



\begin{figure*}[tbh]
\centerline{ \epsfxsize 16cm
   \epsfbox{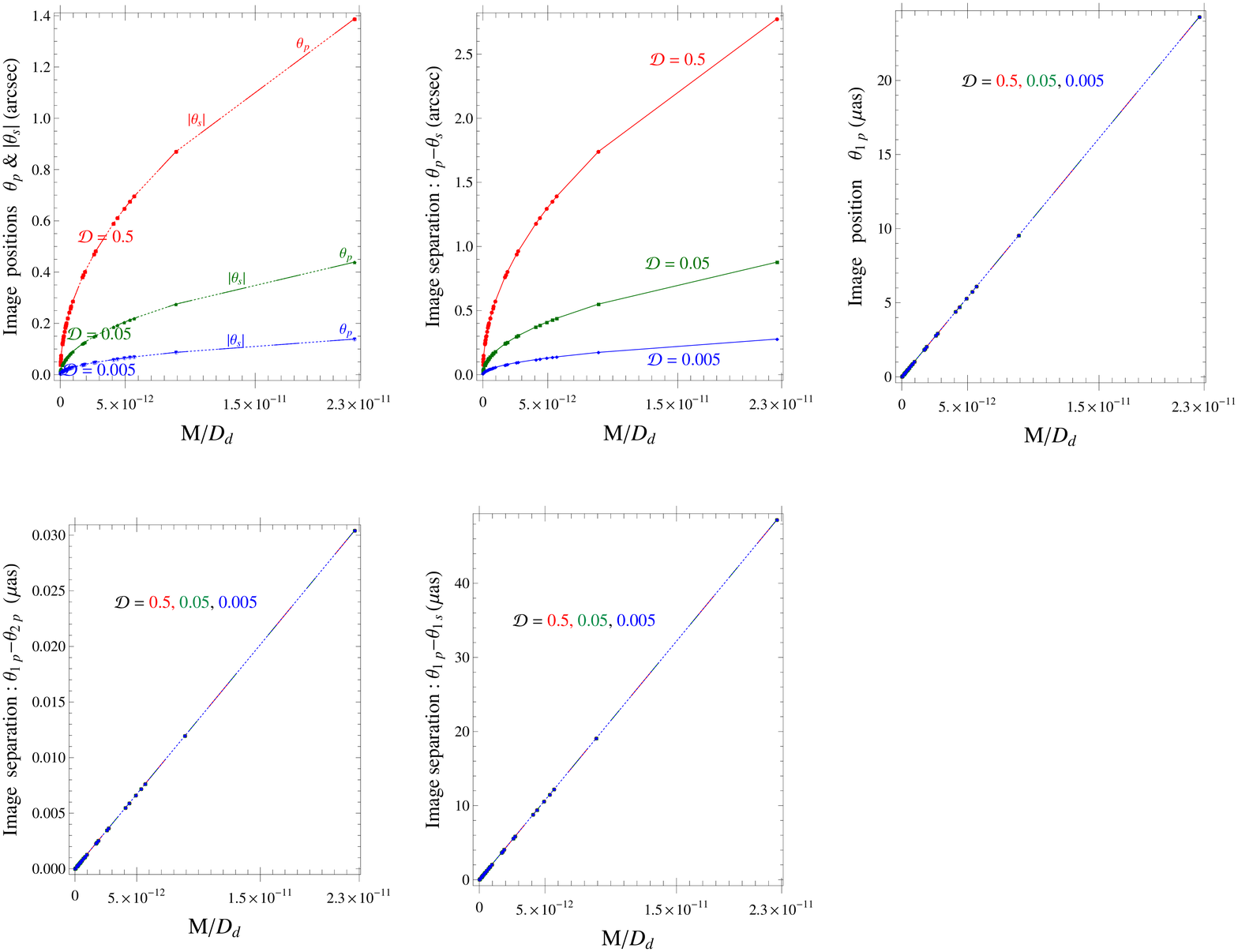}}
 \caption[ ] {(color online). 
{\em Top left and middle}:  The  angular  positions of primary images $\theta_p$ (represented by continuous curves), secondary images $|\theta_s|$
(represented by dotted curves), and their separations  $\theta_p-\theta_s$ are plotted
 against  $M/D_d$ of MDOs at centers of many galaxies for ${\cal D}=0.5,0.05$ and $0.005$. {\em Top right}: The angular
 positions of the relativistic images (on the same side  as the primary image) of the first order $\theta_{1p}$ 
 are plotted against $M/D_d$ for the same values of ${\cal D}$ as in  
figures on left.  The curves for  $\theta_{1p}$ for different values of  ${\cal D}$
intersect for  $(M/D_d)_{1c} \approx 9.31854 \times 10^{-13}$. 
{\em Below middle and right}:  The angular separations $\theta_{1p}-\theta_{2p}$ and $\theta_{1p}-\theta_{1s}$ among relativistic images vs
$M/D_d$ are plotted for the same values of ${\cal D}$ as in figures on top. $\theta_{2p}$ and  $\theta_{1s}$ stand for angular positions of
relativistic images of second order on the primary image side and of first order on the secondary image side, respectively.
The angular source position $\beta = 1 \mu as$ for all  figures. 
}
\label{fig8}
\end{figure*}


\begin{figure*}[tbh]
\centerline{ \epsfxsize 16cm
   \epsfbox{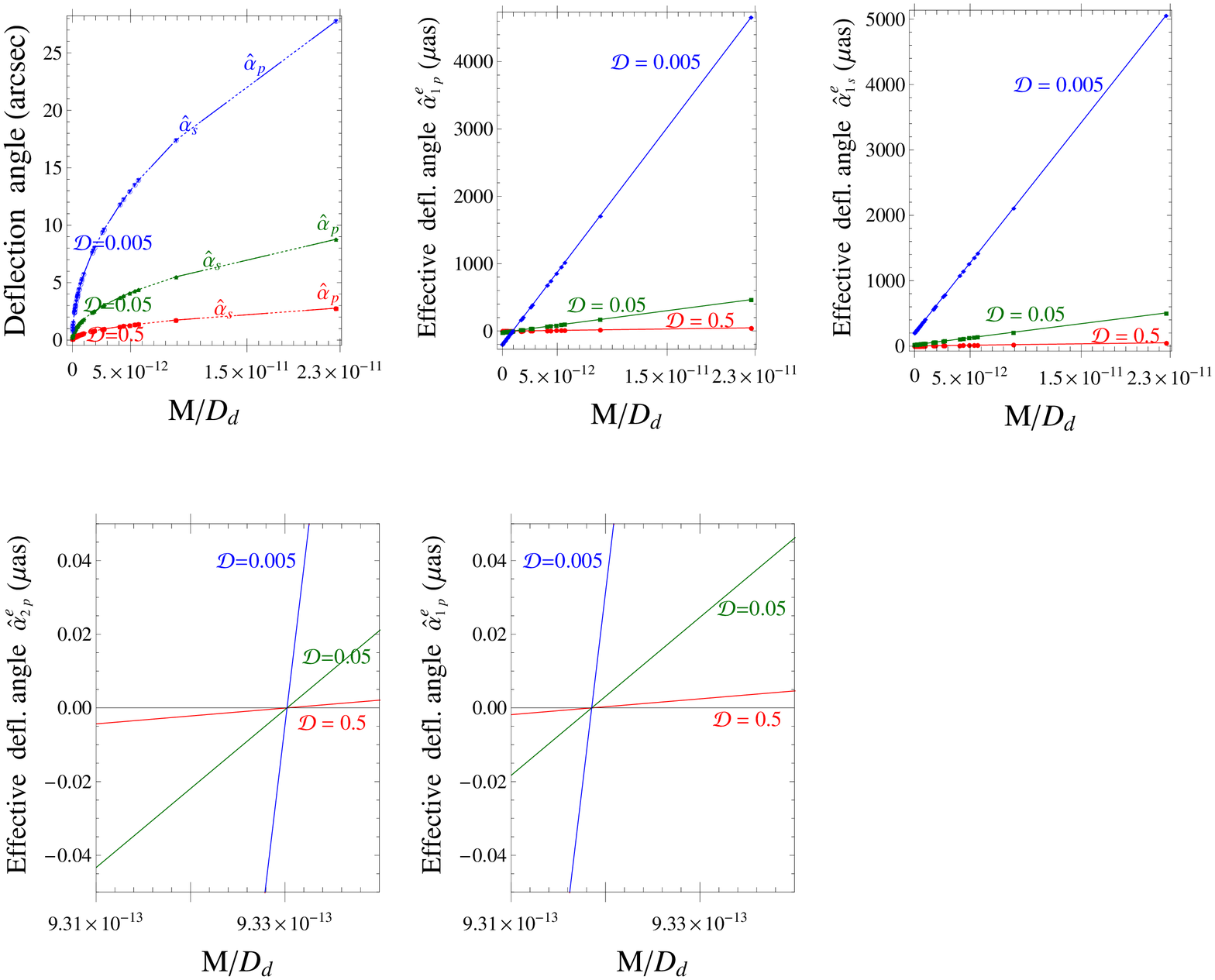}}
 \caption[ ]
{(color online).   
{\em Top left}: The  deflection angles of primary images $\hat{\alpha}_p$ (represented by continuous curves) and of secondary
 images  $\hat{\alpha}_s$  (represented by dotted curves)  are plotted against $M/D_d$ of MDOs at centers of many galaxies 
  for ${\cal D}=0.5,0.05$ and $0.005$. 
{\em Top middle and right}:  The effective deflection angles of  relativistic images  of 
the first order  on the same side as the primary image $\hat{\alpha}^e_{1p}$  and  on the secondary image side  $\hat{\alpha}^e_{1s}$ 
are plotted against $M/D_d$ of MDOs for the same values of ${\cal D}$ as in the left figure. 
{\em Below}: The effective deflection angles of relativistic images on the primary image  side of the second order $\hat{\alpha}^e_{2p}$
and of the first order $\hat{\alpha}^e_{1p}$ vs $M/D_d$ are plotted in the vicinity of zero effective deflection angle for the same 
values of ${\cal D}$.
The curves for different values of ${\cal D}$ on left and right figures  intersect, respectively, 
for  $(M/D_d)_{2c} \approx 9.33022  \times 10^{-13}$ 
and $(M/D_d)_{1c} \approx 9.31854  \times 10^{-13}$. The angular source position $\beta = 1 \mu as$ 
for all figures.
}
\label{fig9}
\end{figure*}

\begin{figure*}[tbh]
\centerline{ \epsfxsize 16cm
   \epsfbox{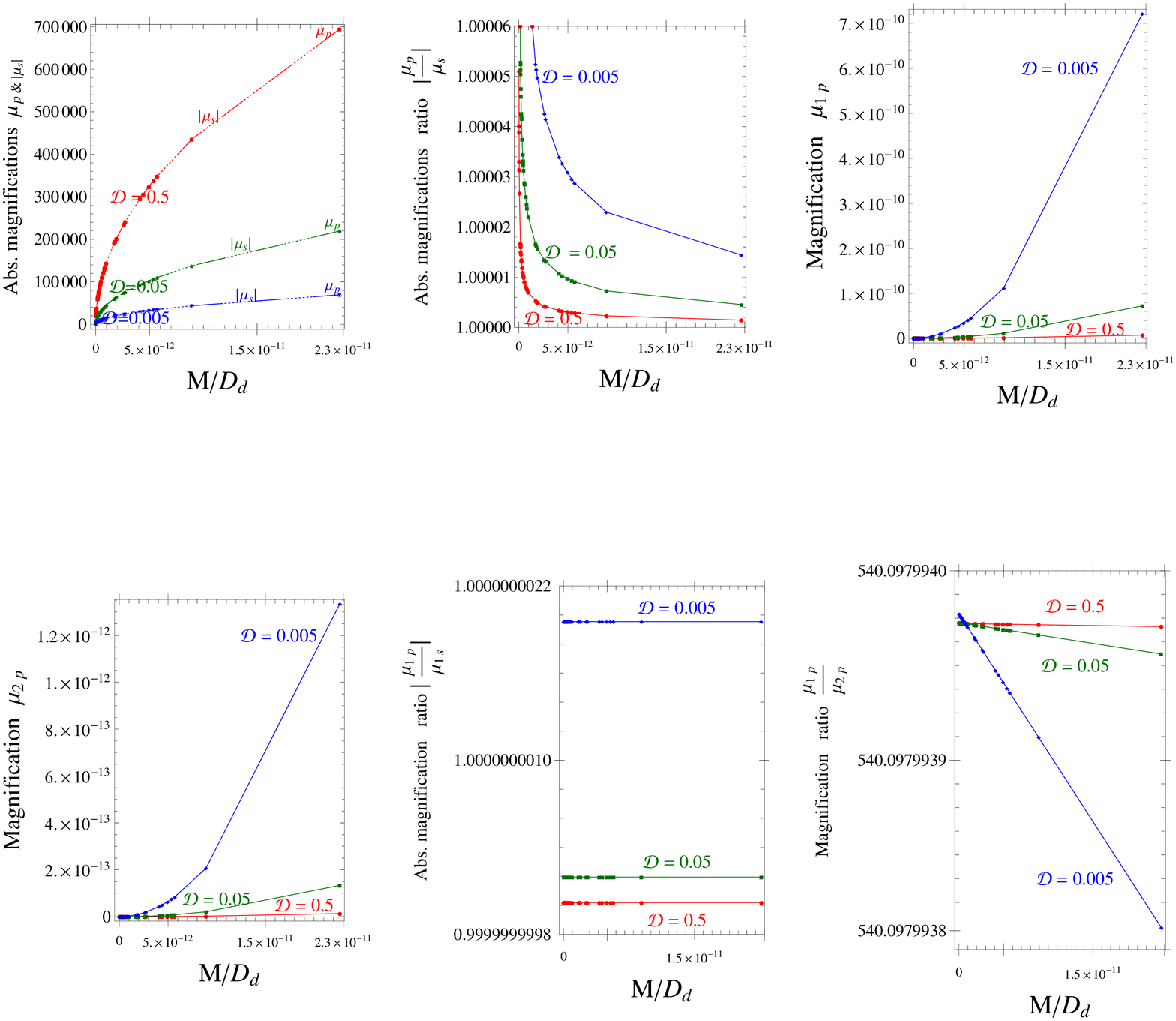}}
 \caption[ ]
{ (color online). 
{\em Top left and top middle}: The  magnifications of primary images $\mu_p$ (represented by continuous curves) and the absolute magnifications of secondary
 images  $|\mu_s|$  (represented by dotted curves), and their ratios $|\mu_p/\mu_s|$   vs $M/D_d$ of MDOs at centers of many galaxies are plotted
for ${\cal D}=0.5,0.05$ and $0.005$.
{\em Top right and below left}:  The magnifications of   relativistic images (on the same side as
 the primary image) of the first order  $\mu_{1p}$  and of  the second order  $\mu_{2p}$  are plotted against $M/D_d$ of MDOs for  same values of ${\cal D}$.
{\em Below middle}: The ratio of absolute magnifications of relativistic images of first order on  the primary image  side $\mu_{1p}$ to
that on the secondary image side $|\mu_{1s}|$ are plotted against $M/D_d$ for ${\cal D}=0.5,0.05$ and $0.005$.
{\em Below right}: The ratio of absolute magnifications of relativistic images (on the primary image side) of the first order $\mu_{1p}$ 
to the second order $\mu_{2p}$ are plotted against $M/D_d$ for same values of ${\cal D}$ as in other plots. The curves  $\mu_{1p}/\mu_{2p}$ vs 
$M/D_d$ for different
values of ${\cal D}$ intersect for  $M/D_d \approx 7.07130 \times 10^{-13}$. For all figures, the angular source position $\beta = 1 \mu as$.
}
\label{fig10}
\end{figure*}


\begin{figure*}[tbh]
\centerline{ \epsfxsize 16cm
   \epsfbox{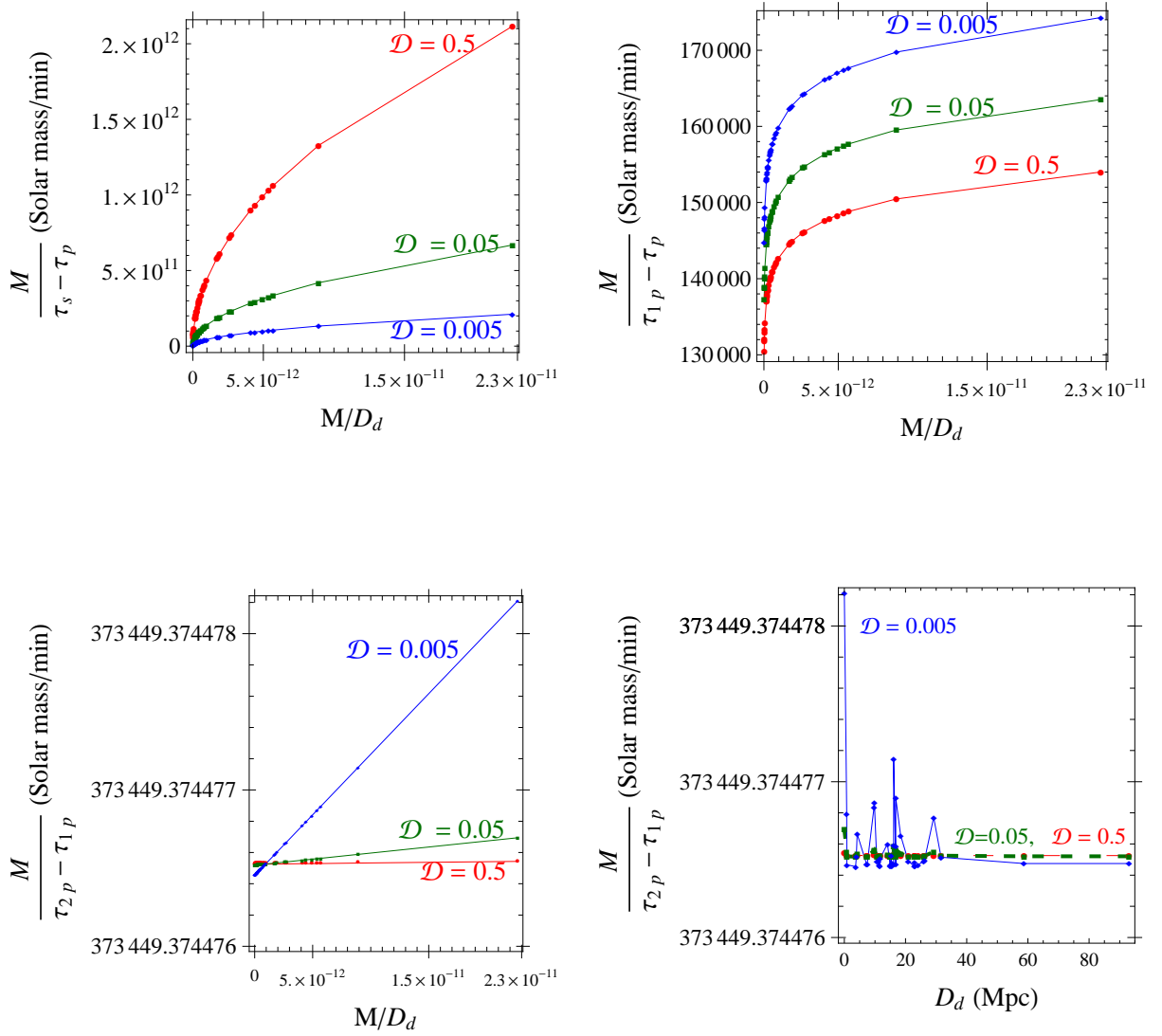}}
 \caption[ ]
{(color online).   {\em Top left, top right, and below left}:
The  ratios  of mass $M$ of the lens  to  the differential time delays of  images
vs $M/D_d$ (the ratio of the mass of the lens to its distance) are plotted  for ${\cal D}=0.5,0.05$ and $0.005$.
$\tau_p$ and $\tau_s$ stand, respectively, for time delays of primary and secondary images, whereas
$\tau_{1p}$ and $\tau_{2p}$, respectively, represent time delays of relativistic images (on the
same side as the primary image) of orders $1$ and $2$. The curves on below left plot intersect for
$M/D_d \approx 9.32438 \times 10^{-13}$. For a given value of ${\cal D}$,
ratios of mass to differential time delay are  strictly increasing function of $M/D_d$.
{\em Below right}:
The  ratios  of mass $M$ of the lens  to  the differential time delay $(\tau_{2p}-\tau_{1p})$ of  images
are plotted against the distance $D_d$ of the lens for same values of ${\cal D}$. The ratio 
$M/(\tau_{2p}-\tau_{1p})$ is not a function of $D_d$. The angular source position $\beta = 1\  \mu as$ for
all   figures.
}
\label{fig11}
\end{figure*}


\begin{figure*}[tbh]
\centerline{ \epsfxsize 12cm
   \epsfbox{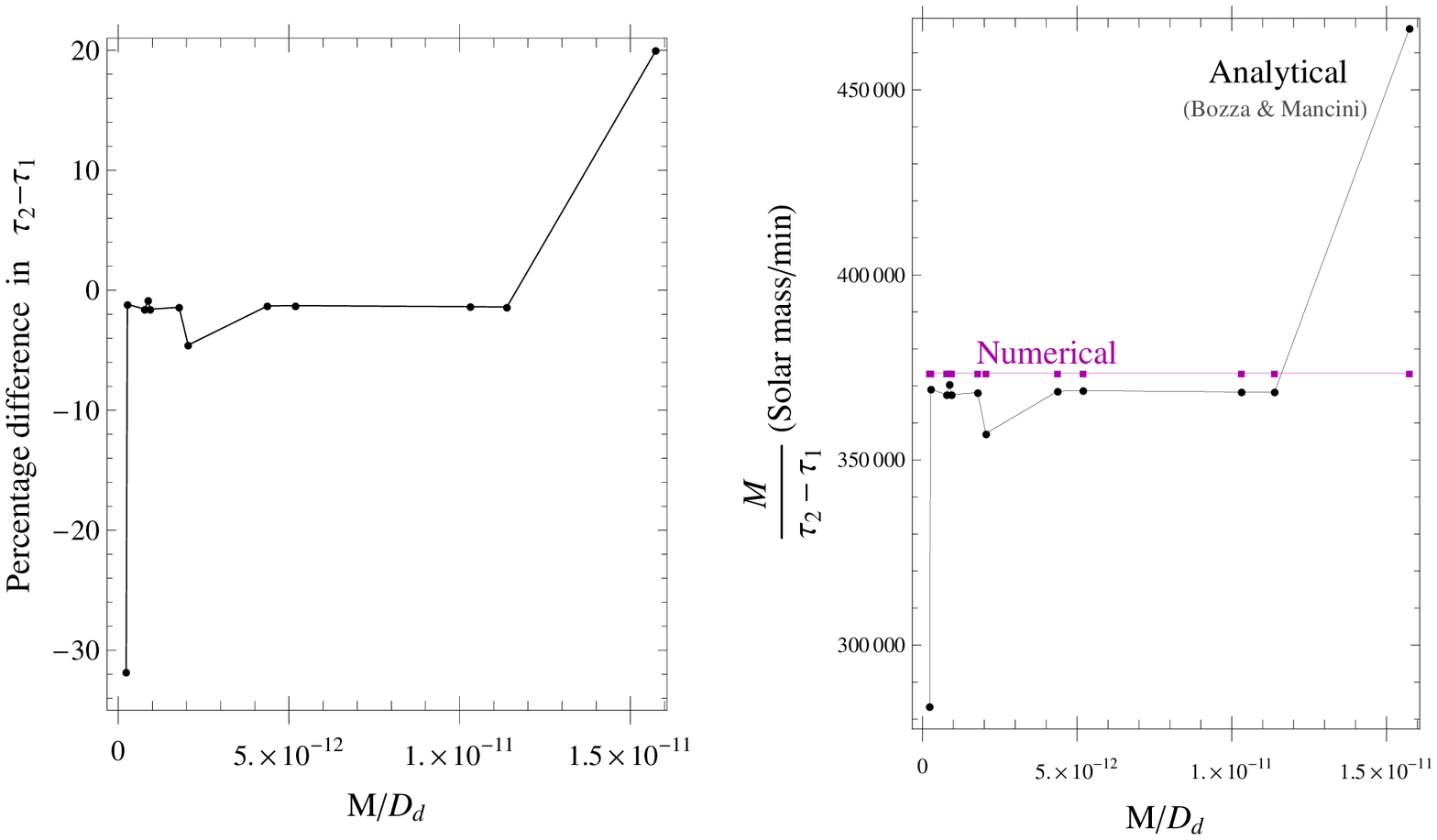}}
 \caption[ ]{(color online). 
{\em Left}: 
The percentage difference in values for differential time delays $(\tau_{2}-\tau_{1})$  obtained by
Bozza and Mancini \cite{BozMan04} (using analytical method) and by us (using numerical method) is plotted against
$M/D_d$  of MDOs at centers of a few galaxies.
$\tau_{1}$ and $\tau_{2}$, respectively, stand for  time delays of relativistic images  of orders $1$ and $2$.
{\em Right}:
The  ratios  of mass $M$ of the lens  to  the differential time delay $(\tau_{2}-\tau_{1})$ 
vs $M/D_d$  are  plotted for both cases.
Our numerical results show that the ratio  $M/(\tau_{2}-\tau_{1})$ is strictly increasing function of $M/D_d$.
Masses and distances of MDOs are taken from  \cite{BozMan04} and are given in Table VII in the Appendix. ${\cal D} = 0.5$ and
the angular source position $\beta=0$ in both figures.
}  
\label{fig12}
\end{figure*}



\section{\label{sec:ManyGalLenses} Gravitational lensing by  MDOs  at centers of many   galaxies}

In this section, we model MDOs at centers of $40$   galaxies as Schwarzschild black hole lenses and, like in the
previous section, study point source GL by them. 
Gebhardt \cite{Geb04} tabulated updated values of masses and distances of
many  MDOs. In Table 4, we consider most of those and arrange in the decreasing order of  $M/D_d$ (i.e., the ratio of mass
$M$ and  the distance $D_d$) of MDOs. (Only for the Galactic MDO, we use the updated values of mass and
distance given in \cite{Eisenetal05}.)
The aim of this section is to study variations in angular positions of images and their separations,
deflection angles (effective deflection angles for relativistic images), magnifications and their ratios, time delays, differential
time delays, and ratios of the mass of the lens to differential time delays due to changes in the value of $M/D_d$ for
${\cal D} \equiv D_{ds}/D_s = 0.5, 0.05$, and $0.005$. We mentioned in the first section  that the central thread
in this paper is the study of relativistic images and we know that these images may be observed only when the lens components
(the observer, the lens, and the source) are perfectly or highly aligned. In view of this we take the angular source 
position $\beta = 1  \mu as$ for computations. (As we also want to  compute magnifications of images due to point source GL,
we do not take $\beta = 0$.) For $\beta = 1 \mu as$,  ${\cal D} = 0.5$, and different values of $M/D_d$ for several MDOs, we
numerically solve the lens equation to obtain image positions of primary and secondary as well as relativistic images 
 of orders $1$ and $2$. 
Further, we compute deflection angles, magnifications, time delays, and differential time delays for primary and secondary as well
as relativistic images.
From deflection angles of relativistic images, we  compute effective deflection angles. In Tables $5$ and $6$, we
present results, respectively, for primary and secondary images, and for relativistic images. Though we do not display results
for deflection angles of primary and secondary images, and effective deflection angles for relativistic images
in tables, we use those in figures. Further, we repeat the entire computations for  ${\cal D} = 0.05$ and $0.005$.
As in the previous section, we do not take either weak or strong gravitational field approximation in any part
of our computations and therefore our results are very accurate. With all results available, we present several plots
and discuss these in the following paragraphs. We do not present some results for relativistic
images on  the secondary image side, because  computations
show that those  results for the same order relativistic images on both sides of
the optical axis are extremely close and therefore graphs for those do not appear resolved on figure.

In Fig. 8, we first plot the angular positions of primary and secondary images, and their separations
against the ratios of the mass of the lens 
to its distance (i.e., $M/D_d$)  for
${\cal D}=0.5, 0.05$, and $0.005$. As expected from well-known analytical expressions for primary and secondary image positions,
the angular positions of these images, for a given value of the angular source position $\beta$, increase with increase
in the values of $M/D_d$ and ${\cal D}$. As we have taken $\beta = 1 \mu as$ (a very small value), the curves for primary
and secondary images are too close to appear resolved on the figure. However, angular position of a primary image is always greater
than that of a secondary image. The angular separation between primary and secondary images increases with 
increase in the value of $M/D_d$ and ${\cal D}$.
 Further, we plot position of the first order relativistic
image $\theta_{1p}$ against $M/D_d$ for ${\cal D}=0.5, 0.05$, and $0.005$. (We do not plot image positions for the second order
relativistic images $\theta_{2p}$, because on chosen scales these do not appear separate from the curve for $\theta_{1p}$.)
For a fixed value of ${\cal D}$ and order $n$, $\theta_{np}$ increases with increase in $M/D_d$. The curves for different values of
${\cal D}$ are too close to appear separate on the figure. Note that there is a critical value of $M/D_d$ at which curves
for different values of ${\cal D}$ intersect. For the first order relativistic images on the primary image side, $(M/D_d)_{1c} 
\approx 9.31854 \times 10^{-13}$.
For $M/D_d < (M/D_d)_{1c}$ and  $M/D_d > (M/D_d)_{1c}$, the value of $\theta_{1p}$ are, respectively, lower  and higher for
higher value of ${\cal D}$. Obviously, for $M/D_d = (M/D_d)_{1c}$, $\theta_{1p}$ is the same for any value of ${\cal D}$.
Similarly, the critical value for the second order relativistic images, $(M/D_d)_{2c} \approx 9.33022  \times 10^{-13}$ and the
above results apply for  $\theta_{2p}$ also. We find that  $(M/D_d)_{2c} >  (M/D_d)_{1c}$. 
For any given value of $M/D_d$ and order $n$, the angular positions of relativistic images (on the secondary image side)
are higher for higher ${\cal D}$. In the next paragraph, we  
show that at critical values for $M/D_d$, the effective deflection angles of relativistic images on the primary side are zero.
The critical value of $M/D_d$ for any give order of relativistic image depends on the value of the angular source position.
We further plot angular separations between first order relativistic images (i.e., $\theta_{1p}-\theta_{1s}$) and between
relativistic images (both on the primary image side) of the first and second orders (i.e., $\theta_{1p}-\theta_{2p}$) against
$M/D_d$ for ${\cal D}=0.5, 0.05$, and $0.005$. Our  results show that variations in angular positions of relativistic images and their
separations are  extremely small (insignificant) due to change in the value of ${\cal D}$.

In Fig. 9, we first plot deflection angles for primary and secondary images against $M/D_d$ (the ratio of mass to distance
of lens) for ${\cal D}=0.5, 0.05$, and $0.005$. As we have  taken the angular source position $\beta$ a very small value, the curves for primary and secondary
images do not appear separate on the figure. The deflection angles for these images increase with increase in the value
of $M/D_d$ (for a fixed value of ${\cal D}$) and decrease with an increase in ${\cal D}$ (for a fixed value of $M/D_d$).
We  now plot the effective deflection angles of the first order relativistic images on each side of the optical axis. The dependence
of these effective deflection angles on ${\cal D}$ and $M/D_d$ is qualitatively similar as for deflection angles for primary
and secondary images. However, there is an important difference. The deflection angles of primary ans secondary images, and
effective deflection angles of relativistic images on the secondary side are always positive. However, the effective deflection 
angles of relativistic images on the primary image side  are negative, zero, or positive depending on the value of $M/D_d$.
The critical value of  the ratio (where the effective deflection angles are zero) for the first and second order relativistic 
images are $(M/D_d)_{1c} \approx  9.31854 \times 10^{-13} $ and  $(M/D_d)_{2c} \approx  9.33022  \times 10^{-13}$ ,
respectively. These values are exactly the same as we obtained
for intersections of curves for image positions for different values of ${\cal D}$ (see the previous paragraph). 

In Fig. 10, we first plot (absolute) magnifications of primary and secondary images, and their ratios against
$M/D_d$ for ${\cal D} = 0.5, 0.05$, and $0.005$. The magnifications increase with increase in the value of
$M/D_d$ (for a fixed valued of ${\cal D}$) as well as  ${\cal D}$ (for a fixed value of 
$M/D_d$). As the chosen angular source position is very small, the ratio of these magnifications is very close 
to $1$. We then plot magnifications of relativistic images of first and second orders (both on the primary 
image side)  for ${\cal D}=0.5, 0.05$, and $0.005$. 
We also plot the ratios of (absolute) magnifications of relativistic images of order $1$ on the primary image
side to the secondary image side against $M/D_d$. We finally plot the ratio of magnifications of relativistic images of orders $1$ and
$2$ (both on the primary image side) vs $M/D_d$ for same values of ${\cal D}$. Compared to the ratios of (absolute) 
magnifications of primary to secondary images,  the ratios of (absolute) magnifications of relativistic images are 
much less sensitive to changes in $M/D_d$ and ${\cal D}$. The curves $\mu_{1p}/\mu_{2p}$ vs $M/D_d$, for different 
values of ${\cal D}$, intersect for $M/D_d \approx 7.07130 \times 10^{-13}$. For $M/D_d$ less and more than its value
on the intersection point, $\mu_{1p}/\mu_{2p}$ are, respectively, higher and lower for a lower value of ${\cal D}$.
As for the case of primary and secondary  images for any given value of ${\cal D}$, magnifications 
of relativistic images increase with increase in the value of $M/D_d$. However, there is a 
substantially and observationally very important difference in both cases. For a given value of $M/D_d$, just
opposite to the case of primary and secondary images, the (absolute) magnifications of relativistic images 
increase with decrease in the value of ${\cal D}$. Therefore, sources nearer to the lens (with 
other conditions remaining the same) would give relativistic images of higher magnifications.

In Fig. 11, we first  study the variation in the ratio of mass of lens to differential time delays among images for
the change in the value of $M/D_d$ for ${\cal D} = 0.5, 0.05$, and $0.005$. We consider differential time 
delays between  secondary and primary images ($\tau_s - \tau_p$), the first order relativistic image on the
primary image side and the primary image ($\tau_{1p}-\tau_p$), and second order and  first order 
relativistic images both on the primary image side ($\tau_{2p} - \tau_{1p}$). We take the angular source position 
$\beta = 1\mu as$. 
The ratio $M/(\tau_s - \tau_p) $ increases with increase in $M/D_d$ (for a fixed value of
${\cal D}$) as well as increase in ${\cal D}$ (for a fixed value of $M/D_d$). It is obvious from the figure that this 
ratio is very sensitive to changes in distances involved in GL. The ratio
 $M/(\tau_{1p} - \tau_{p})$ increases with increase in $M/D_d$ (for a fixed value of ${\cal D}$); however it decreases 
with increase in ${\cal D}$ (for a fixed value of $M/D_d$). This ratio is less (compared to the first we discussed) 
sensitive to changes in $M/D_d$ and ${\cal D}$. The third ratio $M/(\tau_{2p} - \tau_{1p})$ also increases  with increase in 
$M/D_d$ (for a fixed value of ${\cal D}$); however, for a fixed $M/D_d$, the variation against ${\cal D}$ is more fascinating.
There exists a critical value of $M/D_d \approx 9.32438 \times 10^{-13}$ for which curves for different values of ${\cal D}$ 
intersect. For
$M/D_d$ less and more than this critical value, the ratio $M/(\tau_{2p} - \tau_{1p})$ is respectively, less and more for smaller 
value of  ${\cal D}$. This ratio is extremely insensitive to changes in  ${\cal D}$ and $M/D_d$, and  also in $\beta$ (as shown in
Section V). Therefore, this ratio  can be considered 
almost a constant  and can be used to estimate
very accurate values for masses of lenses once the differential time delays $\tau_{2p} - \tau_{1p}$ are known. 
 We also compute this ratio for all MDOs (listed on Table IV) for a few values
of ${\cal D} > 0.5$ (these results are not shown on the plot). We find that the slope at any point on $M/(\tau_{2p} - \tau_{1p})$ vs $M/D_d$ curve tends to $0$ as
${\cal D} \rightarrow 1$.
We finally plot $M/(\tau_{2p} - \tau_{1p})$ against $D_d$ for different values of ${\cal D}$. We find that the ratio
$M/(\tau_{2p} - \tau_{1p})$ is extremely insensitive to changes in ${\cal D}$ as well as $D_d$. Fluctuation in the value
of this ratio decreases with increase in ${\cal D}$.
For a fixed value of
${\cal D}$,  we find that the ratio $M/(\tau_{2p} - \tau_{1p})$  has more than $1$ value for the same value of $D_d$; 
therefore,
this ratio is not a function of $D_d$. However, note that, for any fixed value of ${\cal D}$,   ratios of mass of the lens to the 
differential time delay are strictly increasing functions of $M/D_d$.

\subsection{\label{sec:GalLenses} Comparison with Bozza and Mancini's  results}

Bozza and Mancini (BM) \cite{BozMan04} obtained differential time delays among relativistic images due to GL by a general static
spherically symmetric spacetime. They further modeled MDOs of $12$ galaxies as Schwarzschild lenses, considered ${\cal D} = 0.5$ (i.e., 
the lenses to be symmetrically  situated
between sources and observers) and the angular source position $\beta = 0$. Using their analytical expression
for differential time delays among relativistic images, they computed differential 
time delays between relativistic images of orders $1$ and $2$; i.e., ($\tau_2-\tau_1$), where
$\tau_1$ and $\tau_2$ are, respectively, time delays of relativistic images of orders
$1$ and $2$ for   $\beta=0$.
We put their results in Table $7$ in  decreasing order of $M/D_d$ of lenses. We now consider the
same set of  MDOs and use the same values for mass $M$ and distance 
$D_d$  used in their paper. We do not use the updated values for $M$ and $D_d$ in this
subsection, because we want to compare BM's results with ours.
Our approach is numerical and we do not take either 
weak or strong field approximation. As considered by those authors, we also take ${\cal D} = 0.5$ and $\beta = 0$, and
compute differential time delays of relativistic images of orders $1$ and $2$.
We then compute percentage difference   = $ 100 (x-y)/x $ between ours and their results,  
where $x$ and $y$ are, respectively, differential time delays obtained by us and BM. 
We find that the percentage difference ranges approximately from $-31.8 \%$ to
$20.0 \%$, which are large. It is possible that MDOs of other galaxies (i.e., excluding those considered by BM) give
even higher percentage differences.
We plot the percentage difference in these results against $M/D_d$ (see Fig.~12).
Further,  using BM as well as our results, we compute $M/(\tau_2 - \tau_1)$ for  MDOs. We give these results  in Table $7$.
We then plot this ratio against $M/D_d$ (see Fig.~12). Results obtained by  BM show that the ratio 
$M/(\tau_2 - \tau_1)$  fluctuates quite irregularly (showing no rhythm) with an increase in the value of $M/D_d$ for otherwise constant situation.
There seems to be no physical argument in support for this. 
On the other hand, our results show that $M/(\tau_2 - \tau_1)$ is a strictly increasing (though the increase rate is extremely small)
function of $M/D_d$. As we did not take either weak or strong field approximation at any stage of computation and  performed
numerical computation with  high precision, our results are  very accurate. Therefore, we consider percentage 
differences in results as percentage errors in their results.

\section{\label{sec: Discussion and Summary}  Discussion and Summary}

It is well-known that the observation of primary and secondary images due to GL by an MDO 
at the center of a galaxy is very difficult due to a large extinction of electromagnetic radiation 
(larger extinction for smaller wavelength) in the vicinity of a galactic center. In addition,
radiations at several wavelengths from materials accreting on an MDO badly hinders observation of 
these images. These obstacles would be even bigger for relativistic images as, compared to 
primary and secondary images, these are formed much closer to the center of a galaxy. Unfortunately,
 observations of relativistic images would be much more difficult due to  some additional reasons. 
Relativistic images are very much demagnified, unless the
lens components (the source, the lens, and the observer) are perfectly or highly aligned
($\beta << 1 \mu as$), and therefore these images are extremely difficult to be observed. Supernovae
could be more suitable sources for observation of relativistic images, but
the probability that a supernova  will be highly aligned with the lens and observer is 
extremely small. However, there is a silver lining to the demagnification problem associated with 
observation of relativistic images:  magnifications of relativistic images increase rapidly
with the decrease in the value of ${\cal D }$; i.e., with the decrease in  the source-lens 
distance for otherwise constant situation. Thus, sources closer to a galactic center would give less 
demagnified  relativistic images. Despite this, there is no doubt that the observation of relativistic 
images would be a Herculean task. However, with improved observational facilities in future and through 
lucky observations (due to a bright source close to a galactic center and highly aligned with the galactic center 
and the observer), the relativistic images could  possibly be detected some day. Today's these almost 
unthinkable events may be tomorrow's observations. The detection of relativistic images would be definitely 
one of the most important discovery in astronomy and would have immense implications for general relativity 
and relativistic astrophysics. For examples, these observations would provide a test for the general theory  of relativity
in a strong gravitational field. In \cite{VE00}, we discussed that observation of 
relativistic images would give upper bound to the compactness of   MDOs and therefore would strongly support 
that these MDOs are black holes. The measurements of physical quantities for relativistic images would
also give very accurate values for masses and distances of black holes.

For any fixed value of ${\cal D}$, the ratio of  mass $M$ of a Schwarzschild lens to  differential time delay 
$(\tau_{2p}-\tau_{1p})$ or $(\tau_{2s}-\tau_{1s})$ is not a function of the lens-observer distance $D_d$; however, it is a strictly
increasing function of $M/Dd$. Computations show that the ratio $M/(\tau_{2p}-\tau_{1p})$ is {\em extremely}
insensitive to changes in the angular source position as well as the observer-lens and lens-source distances, and therefore this awesome physical 
quantity must be treasured as an almost constant for purpose of measurements. Thus, once we succeed in 
 detecting relativistic images and measuring the differential time delay, we can immediately 
compute a very accurate value of mass of the MDO acting as a lens. (The accuracy of the result for the mass of the MDO
will however depend on the accuracy of the measurement of differential time delay.) 
Our computations show that for $M/D_d = (0,2.265 \times 10^{-11})$  and ${\cal D} = [0.005,1)$, 
\begin{equation}
M  \approx 3.734493744773 \times 10^5 \  (\tau_{2p}-\tau_{1p})   , 
\end{equation}
where mass $M$ of the MDO  and the differential time delay $\tau_{2p}-\tau_{1p}$ are expressed in units of {\em solar mass} and {\em minute}, 
respectively.
As relativistic images can  possibly be observed when the 
lens components are perfectly
or highly aligned, we took the angular source position $\beta = 1\mu as$ for computations. However, we found  extremely small changes 
in  results for  computations with
$\beta = 0$. Once, the value of mass of the MDO is known, its distance can 
be computed from  the results given in Fig. 8 (below). Angular separations between relativistic images depend on the ratio
$M/D_d$, but fortunately it is extremely insensitive to the change in the value of ${\cal D}$. This would help us measure
the distance of the MDO very accurately once we have the mass of the MDO and the angular separation between relativistic 
images is measured. It is worth mentioning
that accuracies in determination of  distances of black holes would, however, depend on accuracies of our measurements
of differential time delays and angular  separations between  relativistic images.
The dependence  of (absolute) magnifications ratio  of relativistic images of the first order (i.e., $|\mu_{1p}/\mu_{1s}|$)
on  $M/D_d$ is extremely small (see Fig~10). Therefore, measurement of the magnifications ratio would
give very accurate  value for ${\cal D}$. This result with already obtained value for observer-lens distance $D_d$ would give 
observer-source distance $D_s$.

Effective deflection angles of relativistic images play a very significant role in analyzing and understanding these
images. The deflection angles for primary-secondary image pair as well as relativistic images of Schwarzschild black
hole lensing are always positive. The effective deflection angles of relativistic images of any order on the secondary
image side  are always positive. However, the effective deflection angles of relativistic images
of any order on the primary image side may be positive, zero, or negative depending on the value of the angular source
position $\beta$ and the ratio of mass of the lens to its distance (i.e., $M/D_d$). 
For  a relativistic 
image (on the primary image side) of any order $n$ and for any (nonzero) value of $M/D_d$ of the lens,
there exists a critical angular source position $\beta_{nc}$
such that the effective deflection angle $\alpha^{\hat e}_{np}$ for that relativistic image is zero. For 
$\beta < \beta_{nc}$, $\alpha^{\hat e}_{np} > 0$, and for $\beta > \beta_{nc}$, $\alpha^{\hat e}_{np} < 0$. For a given 
value of $M/D_d$, $\beta_{nc}$ is smaller  for smaller $n$. All sources at $\beta = \beta_{nc}$ are lensed to give rise to 
$n^{th}$  order relativistic images (on primary image side) at  the same angular position  $\theta_{np}= \beta_{nc}$. For a fixed value of 
$M/D_d$, the angular positions of relativistic images are extremely insensitive to changes in the angular source position
as well as the lens-source distance. 
However, for a theoretical interest, it is worth noting that for $\beta < \beta_{nc}$
and $\beta > \beta_{nc}$, the value of $\theta_{np}$ is, respectively,  higher and lower  for higher value of  ${\cal D}$. The 
critical angular source position plays a role of flipping point for image positions with respect to the change in the value
of ${\cal D}$. These results help us conclude the following:
For different sources at the same angular position, relativistic images with positive, zero,
and negative effective deflection angles  have, respectively, bigger, equal, and smaller (absolute) angular positions for 
bigger values of ${\cal D}$.
This is also  true for primary and secondary images, as deflection angles for them are always positive.
These conclusions can be also derived from the lens equation. Therefore, these results support correctness of our
numerical computations.

In GL  observations, differential time delays among images (not the time delays of individual images) have been measured
until now. For this reason, studies of time delays of images have not drawn enough attention. The most well-known 
book on GL gives an expression for time delays of gravitationally lensed images [see Eq. (4.67) in \cite{book1}]. The 
equation has an additive constant term. The authors clearly stated that the constant term is the same for all rays from the
observer to the source plane. Though the value for the constant term is not yet determined, the expression 
given in the book is good enough to evaluate differential time delays among images. This is because the constant term cancels for images 
of the same source by the same lens.
Interestingly, Rafikov and Lai \cite{RL06} have recently pointed out that time delays of individual images are in fact
measurable. This motivated us to first compute time delays of images and then use these results to compute differential
time delays among them. As Eq. (4.67) in \cite{book1} cannot be used to compute time delays of images, we used the method given in  
Weinberg's book. Our results for time delays of primary and secondary images turn out to be non-intuitive and very fascinating
(see Fig. 6).  For instance, time delays of primary images are always smaller for sources nearer to the lens for otherwise 
constant situation.  Time delays results for relativistic images are also very interesting and important.

GL as well as gravitational retro-lensing  give rise to very much demagnified images
due to light deflections in strong gravitational field. Images due to these   two phenomena can be 
observationally  easily differentiated  by the fact that the images due to the latter are ``orphans" in a sense that 
these are not accompanied by  primary-secondary images pair as  their ``parents".  In this paper, we have studied 
only images due to GL. Eiroa and Torres \cite{ET04} studied retro-lensing by a Schwarzschild black hole. They compared magnifications
of gravitationally retro-lensed images and relativistic images of GL of the same order (i.e., the number of turns
a light ray makes around the lens before reaching the observer) and found that the former is significantly 
greater than the latter. Black holes have angular momentum. Therefore, Cunninghom and Bardeen \cite{CB73} and Rauch and
Blandford \cite{RB94} pioneered Kerr black hole lensing. As there has been mounting observational evidence in 
support of existence of black holes, Kerr lensing has become a very lively research topic (see \cite{VazEst04} and
references therein) these days. It is worth investigating the behavior of the ratio of mass $M$ of the lens to 
differential time delays of images of strong field lensing due to changes in $\beta$, $M/D_d$, ${\cal D}$, and  $a/m$ ($a$ is the 
rotational  parameter of the Kerr metric). These investigations are likely to have immense
implication for relativistic astrophysics.

With increasing observational support for MDOs at centers of galaxies and stellar size  black hole candidates to be 
black holes, the pressure to believe in the existence of black holes in the Universe has began to mount. However, by the definition of a black
hole, there is no and there cannot be an iron-clad observational evidence that a black hole candidate is indeed a black hole.
Given that the {\em weak cosmic censorship hypothesis} (WCCH) of Penrose is still unproven (see \cite{CCH} and references therein),
there is no compelling scientific reason to accept that all black hole candidates are black holes and none of them can be interpreted
as a {\em naked (visible) singularity}. Despite the fact that the concept of naked singularity does not ``smell right"  to majority
of researchers, it may not be wise to completely ignore the possibility of existence of naked singularities.
 Researchers think that in the vicinity of a spacetime singularity, a mysterious violent 
marriage of general relativity and quantum physics is solemnized and opportunities to observe these (through outgoing 
geodesics from there to us)
could help us obtaining an unanimously acceptable viable quantum gravity theory. Philosophically, it is not clear to us why the 
nature should be malicious to always hide such awesome holy marriages from us. Inspired by these ideas, we initiated a new theoretical 
research project using GL phenomena that investigates whether or not black holes and naked singularities could be observationally
differentiated (see \cite{VNC98, VE02, VK08}). Our computations yielded encouraging distinctive results.
Whether or not the weak cosmic censorship hypothesis of Penrose finally turns out to be true, there has to be a cosmic
censorship which forbids arbitrary large values of {\em those} nakedness parameters [e.g.,  $(Q/M)^2$ in the Reissner-Nordstr\"om solution to the Einstein-Maxwell
equations, where $Q$ and $M$ stand for
 electric charge and mass parameters, respectively] which make the system unphysical. Motivated by this idea, we hypothesize 
 a new cosmic censorship: Generically, marginally and strongly naked singularities do not occur in a realistic
gravitational collapse. (For definitions of weakly naked, marginally naked, and strongly naked singularities, see 
\cite{VK08}). The new cosmic censorship hypothesis (CCH) allows the existence of  weakly naked singularities,
but does not say that these do exist. This hypothesis does not
imply that the well-known weak cosmic censorship hypothesis is incorrect. Rather, it says that in case the WCCH of Penrose turns out
to be incorrect, the   new cosmic censorship will  hold good. In \cite{CVE01}, we showed that a Vaidya naked singularity is 
weakly naked and therefore it is not a counter-example  to the new  CCH. The proof of the pudding is in the eating. It may be
of an  astrophysical interest to investigate this subject further. We will  discuss the new CCH
in  detail in \cite{KSV}.

\acknowledgments

Thanks are due to J.~M.~Aguirregabiria, H.~M.~Antia, M.~Batrelmann, and D.~Narasimha for helpful correspondence related to 
verification of  some results of Fig.~6 (top left). Thanks are also due to  F.~Eisenhauer  for giving me the reference \cite{Geb04}.

\section{appendix}

\section*{Tables I-VII}

Tables I through VII are given in this section.

\begingroup
\squeezetable
\begin{table*}
\caption{\label{tab:Table1} Angular  positions, bending angles, magnifications, and time delays
of primary and secondary images due to GL by the Galactic MDO  modeled as a Schwarzschild black hole. $\beta$ stands for the
angular source position. $\theta$, $\hat{\alpha}$, $\mu$, and $\tau$ stand, respectively, for   angular positions,
deflection angles, magnifications, and time delays of images, with $p$ and $s$ subscripts, respectively, for primary and secondary images.
$\tau_s-\tau_p$ stands for the differential time delay of the secondary image with respect to the primary image. All angles
are expressed in {\em arcsec}, and time delays and differential time delays are given in {\em minutes}. 
{ \bf (a)} The Galactic MDO (lens) has  mass $M= 3.61 \times 10^6 M_{\odot}$, which is at distance $D_d =  7.62$ {\em kpc}. 
 $M/D_d \approx 2.26 \times  10^{-11}$, where $M \equiv  MG/c^2$. The ratio of the lens-source distance to the observer-source 
distance ${\cal D}=0.5$.  Results in this table are taken from our recent paper \cite{VK08}.
        }
\begin{ruledtabular}
  \begin{tabular}{l|ccccc|cccc}  
\multicolumn{1}{c|}{$\beta$}&
\multicolumn{5}{c|}{Secondary image}&
\multicolumn{4}{c}{Primary image}\\
&    ${\theta}_s$            &         $\hat{\alpha}_s$ &    $\mu_s$     &   $\tau_s$     & $\tau_{s}-\tau_{p}$  &    $\theta_{p}$    &  $\hat{\alpha}_{p}$ &    $\mu_{p}$   &   $\tau_{p}$    \\
\hline
$0	 $&$  -1.388176   $&$	2.776352  $&$	  \times  $&$	14.9220910 $&$	  0	  $&$   1.388176  $&$   2.776352  $&$	\times   $&$    14.9220910 $\\
$10^{-6} $&$  -1.388176   $&$	2.776353  $&$	-694084.2 $&$	14.9220919 $&$	0.000002  $&$	1.388177  $&$	2.776351  $&$	694085.2 $&$	14.9220902$\\
$10^{-5} $&$  -1.388171	  $&$   2.776362  $&$	-69407.97 $&$	14.9220995 $&$	0.000017  $&$	1.388181  $&$	2.776342  $&$	69408.97 $&$	14.9220825$\\
$10^{-4} $&$  -1.388126	  $&$   2.776452  $&$	-6940.347 $&$	14.9221763 $&$	0.000171  $&$	1.388226  $&$	2.776252  $&$	6941.347 $&$	14.9220057$\\
$10^{-3} $&$  -1.387676	  $&$   2.777353  $&$	-693.5848 $&$	14.9229442 $&$	0.001706  $&$	1.388676  $&$	2.775353  $&$	694.5848 $&$	14.9212382$\\
$10^{-2} $&$  -1.383185	  $&$   2.786370  $&$	-68.90982 $&$	14.9306363 $&$	0.017060  $&$	1.393185  $&$	2.766370  $&$	69.90982 $&$	14.9135764$\\
$10^{-1} $&$  -1.339077	  $&$   2.878153  $&$	-6.454348 $&$	15.0089452 $&$	0.170636  $&$	1.439076  $&$	2.678152  $&$	7.454345 $&$	14.8383092$\\
$1	 $&$  -0.975480	  $&$   3.950960  $&$	-0.322455 $&$	15.9468061 $&$	1.742193  $&$	1.975475  $&$	1.950951  $&$	1.322453 $&$	14.2046135$\\
$2	 $&$  -0.710863	  $&$   5.421726  $&$	-0.073840 $&$	17.3803344 $&$	3.687537  $&$	2.710855  $&$	1.421709  $&$	1.073838 $&$	13.6927977$\\
$3	 $&$  -0.543786	  $&$   7.087573  $&$	-0.024114 $&$	19.2981794 $&$	5.987040  $&$	3.543776  $&$	1.087553  $&$	1.024113 $&$	13.3111391$\\
$4	 $&$  -0.434559	  $&$   8.869117  $&$ 	-0.009696 $&$ 	21.7471848 $&$ 	8.734391  $&$ 	4.434547  $&$ 	0.869094  $&$ 	1.009695 $&$ 	13.0127934$\\
$5	 $&$  -0.359561	  $&$   10.71912  $&$	-0.004521 $&$	24.7549479 $&$	11.98479  $&$	5.359549  $&$	0.719098  $&$	1.004521 $&$	12.7701628$\\
 \end{tabular}
\end{ruledtabular}
 \end{table*}
 \endgroup
\begingroup
\squeezetable
\begin{table*}
\caption{\label{tab:Table2} 
Effective deflection  angles, magnifications, and time delays of relativistic images (on the same side
as the primary image) due to GL by the Galactic MDO  modeled as a Schwarzschild black hole.  $\hat{\alpha}^e$, $\mu$, 
and $\tau$ stand, respectively, for effective deflection angles, magnifications, and time delays, 
with $1p$ and $2p$ subscripts, respectively,  are used for the  first and the second order relativistic images on the same side as 
the primary image. $\tau_{2p}-\tau_{1p}$, $\tau_{1p}-\tau_{p}$, and $\tau_{1p}-\tau_{s}$ are  differential time delays.
The angular positions of the first and second order relativistic images on the primary image side are, respectively,
$\theta_{1p} \approx 24.30283 \ \mu as$ and $\theta_{2p} \approx 24.27240 \ \mu as$ for all values of angular source
position $\beta$ considered in this table.
{\bf (a)} The angular source   positions   and the  effective deflection angles are, respectively, expressed in {\em arcsec} and 
$\mu as$, whereas time delays and
differential time delays are given in {\em minutes}. The mass and distance of the lens  are as given in ${(a)}$ 
of Table I.} 
\begin{ruledtabular}
  \begin{tabular}{l|cccc|ccccc}  
\multicolumn{1}{c|}{$\beta$}&
\multicolumn{4}{c|}{Second order (inner) relativistic image}&
\multicolumn{5}{c} {First order (outer) relativistic image }\\
&       $\hat{\alpha}^{e}_{2p}$ &    $\mu_{2p}$ & $\tau_{2p}$  & $\tau_{2p}-\tau_{1p}$ &  $\hat{\alpha}^{e}_{1p}$ & $\mu_{1p}$ & $\tau_{1p}$  
                                                                                       &   $\tau_{1p}-\tau_p$  & $\tau_{1p}-\tau_s$ \\
\hline
$0	 $&$     48.544793 $&$	   \times	     $&$   48.0274920 $&$	9.666638229 $&$  48.605666 $&$	\times	             $&$        38.3608537 $&$	23.438763 $&$	23.438763$\\
$10^{-6} $&$	 46.544793 $&$	1.34 \times 10^{-14} $&$   48.0274920 $&$	9.666638229 $&$	 46.605666 $&$	7.21 \times 10^{-12} $&$	38.3608537 $&$	23.438764 $&$	23.438762$\\
$10^{-5} $&$	 28.544793 $&$	1.34 \times 10^{-15} $&$   48.0274920 $&$	9.666638229 $&$	 28.605666 $&$	7.21 \times 10^{-13} $&$	38.3608537 $&$	23.438771 $&$	23.438754$\\
$10^{-4} $&$	-151.45521 $&$	1.34 \times 10^{-16} $&$   48.0274920 $&$	9.666638229 $&$	-151.39433 $&$	7.21 \times 10^{-14} $&$	38.3608537 $&$	23.438848 $&$	23.438677$\\
$10^{-3} $&$	-1951.4552 $&$	1.34 \times 10^{-17} $&$   48.0274922 $&$	9.666638229 $&$	-1951.3943 $&$	7.21 \times 10^{-15} $&$	38.3608540 $&$	23.439616 $&$	23.437910$\\
$10^{-2} $&$	-19951.455 $&$	1.34 \times 10^{-18} $&$   48.0275225 $&$	9.666638229 $&$	-19951.394 $&$	7.21 \times 10^{-16} $&$	38.3608843 $&$	23.447308 $&$	23.430248$\\
$10^{-1} $&$	-199951.46 $&$	1.34 \times 10^{-19} $&$   48.0305628 $&$	9.666638231 $&$	-199951.39 $&$	7.21 \times 10^{-17} $&$	38.3639246 $&$	23.525615 $&$	23.354979$\\
$1	 $&$    -1999951.5 $&$	1.34 \times 10^{-20} $&$   48.3347132 $&$	9.666638248 $&$	-1999951.4 $&$ 	7.21 \times 10^{-18} $&$	38.6680749 $&$	24.463461 $&$	22.721269$\\
 \end{tabular}
\end{ruledtabular}
 \end{table*}
 \endgroup

\begingroup
\squeezetable
\begin{table*}
\caption{\label{tab:Table3}  
Effective deflection angles, magnifications, and time delays of relativistic images (on the same side
as the secondary image) due to GL by the Galactic MDO  modeled as a Schwarzschild black hole.  $\hat{\alpha}^e$, $\mu$, 
and $\tau$ stand, respectively, for  effective deflection angles, magnifications, and time delays, 
with subscripts $1s$ and $2s$, respectively, are used for the  first and the second order relativistic images  on the secondary image  side.
$\tau_{1s}-\tau_{1p}$, $\tau_{1s}-\tau_{p}$, $\tau_{1s}-\tau_{s}$ , and $\tau_{2s}-\tau_{1p}$  are
differential time delays.  
The angular positions of the first and second order relativistic images on the secondary image side are, respectively,
$\theta_{1s} \approx -24.30283 \ \mu as$ and $\theta_{2s} \approx -24.27240 \ \mu as$ for all values of angular source
position $\beta$ considered in this table.
${\bf (a)}$ The same as ${(a)}$  of Table II.} 
\begin{ruledtabular}
  \begin{tabular}{l|cccccc|cccc}  
\multicolumn{1}{c|}{$\beta$}&
\multicolumn{6}{c|}{First (outer) relativistic image}&
\multicolumn{4}{c} {Second (inner) relativistic image }\\
&       $\hat{\alpha}^{e}_{1s}$ &    $\mu_{1s}$ & $\tau_{1s}$  & $\tau_{1s}-\tau_{1p}$ & $\tau_{1s}-\tau_p$ & $\tau_{1s}-\tau_{s}$ & 
 
$\hat{\alpha}^{e}_{2s}$ & $\mu_{2s}$ & $\tau_{2s}$  &   $\tau_{2s}-\tau_{1p}$ \\
\hline
 $0        $&$ 48.605666 $&$   \times             $&$ 38.3608537 $&$       0             $&$ 23.438763 $&$ 23.438763 $&$ 48.544793 $&$     \times       $&$ 48.0274920 $&$ 9.666638229 $\\
 $10^{-6}  $&$ 50.605666 $&$ -7.21\times 10^{-12} $&$ 38.3608537 $&$ 2.99\times 10^{-11} $&$ 23.438764 $&$ 23.438762 $&$ 50.544793 $&$ -1.34\times 10^{-14} $&$ 48.0274920 $&$ 9.666638229 $\\
 $10^{-5}  $&$ 68.605666 $&$ -7.21\times 10^{-13} $&$ 38.3608537 $&$ 2.99\times 10^{-10} $&$ 23.438771 $&$ 23.438754 $&$ 68.544793 $&$ -1.34\times 10^{-15} $&$ 48.0274920 $&$ 9.666638230 $\\
 $10^{-4}  $&$ 248.60567 $&$ -7.21\times 10^{-14} $&$ 38.3608537 $&$ 2.99\times 10^{-9}  $&$ 23.438848 $&$ 23.438677 $&$ 248.54479 $&$ -1.34\times 10^{-16} $&$ 48.0274920 $&$ 9.666638232 $\\
 $10^{-3}  $&$ 2048.6057 $&$ -7.21\times 10^{-15} $&$ 38.3608540 $&$ 2.99\times 10^{-8}  $&$ 23.439616 $&$ 23.437910 $&$ 2048.5448 $&$ -1.34\times 10^{-17} $&$ 48.0274923 $&$ 9.666638259 $\\
 $10^{-2}  $&$ 20048.606 $&$ -7.21\times 10^{-16} $&$ 38.3608846 $&$ 2.99\times 10^{-7}  $&$ 23.447308 $&$ 23.430248 $&$ 20048.545 $&$ -1.34\times 10^{-18} $&$ 48.0275228 $&$ 9.666638528 $\\
 $10^{-1}  $&$ 200048.61 $&$ -7.21\times 10^{-17} $&$ 38.3639276 $&$ 2.99\times 10^{-6}  $&$ 23.525618 $&$ 23.354982 $&$ 200048.54 $&$ -1.34\times 10^{-19} $&$ 48.0305658 $&$ 9.666641214 $\\
 $1        $&$ 2000048.6 $&$ -7.21\times 10^{-18} $&$ 38.6681048 $&$ 2.99\times 10^{-5}  $&$ 24.463491 $&$ 22.721299 $&$ 2000048.5$&$  -1.34\times 10^{-20} $&$ 48.3347430 $&$ 9.666668077 $\\
 \end{tabular}
\end{ruledtabular}
 \end{table*}
 \endgroup


\begingroup
\squeezetable
\begin{table*}
\caption{\label{tab:Table4}  Masses and  distances of MDOs at centers of  $40$ galaxies are  presented in the decreasing
order of  dimensionless ratio of mass to  distance [i.e., $M/D_d \equiv MG/(c^2 D_d)$] of MDOs. The mass and distance of the Galactic MDO and all other MDOs are,
respectively, taken from \cite{Eisenetal05} and \cite{Geb04}.
      } 
\begin{ruledtabular}
  \begin{tabular}{lccc|lccc}  

$\text{MDO in}$ &  $\text{Mass} ~M      $  & $\text{Distance} ~D_d$ & $ \frac{M}{D_d}$  & $\text{MDO in}$ &  $\text{Mass} ~M      $  & $\text{Distance} ~D_d$ &  $ \frac{M}{D_d}\vspace*{-0.04in}$  \\
$\text{~galaxy}$ &  $\text{in} ~M_{\odot}$  & $\text{in} ~Mpc$       & $              $  & $\text{~galaxy}$ &  $\text{in} ~M_{\odot}$  & $\text{in} ~Mpc      $  & $                              $ \\

\hline
$\text{Milky Way}    $&$        3.61\times10^6$&$  	0.00762$&$	2.26467\times10^{-11} $&$  \text{NGC5845}    $&$	        2.4\times10^8 $&$	        25.9 $&$	4.42959\times10^{-13}$\\
$\text{NGC4486(M87)} $&$	3.0\times10^9 $&$   	16.1   $&$	8.90733\times10^{-12} $&$  \text{NGC3377}    $&$	        1.0\times10^8 $&$	        11.2 $&$	4.26810\times10^{-13}$\\
$\text{NGC4649}      $&$	2.0\times10^9 $&$   	16.8   $&$	5.69080\times10^{-12} $&$  \text{NGC3608}    $&$  	        1.9\times10^8 $&$	        22.9 $&$	3.96616\times10^{-13}$\\
$\text{NGC4594}      $&$        1.1\times10^9 $&$	9.80   $&$	5.36561\times10^{-12} $&$  \text{NGC4473}    $&$	        1.1\times10^8 $&$	        15.7 $&$	3.34923\times10^{-13}$\\
$\text{NGC3115}	     $&$        1.0\times10^9 $&$	9.70   $&$	4.92811\times10^{-12} $&$  \text{NGC6251}    $&$	        5.3\times10^8 $&$	        93.0 $&$	2.72424\times10^{-13}$\\
$\text{NGC224(M31)}  $&$	7.0\times10^7 $&$	0.760  $&$	4.40288\times10^{-12} $&$  \text{NGC7052}    $&$	        3.3\times10^8 $&$	        58.7 $&$	2.68737\times10^{-13}$\\
$\text{IC1459}	     $&$        2.5\times10^9 $&$   	29.2   $&$	4.09270\times10^{-12} $&$  \text{NGC2787}    $&$  	        4.1\times10^7 $&$	        7.50 $&$	2.61321\times10^{-13}$\\
$\text{NGC5128(cenA)}$&$	2.4\times10^8 $&$	4.20   $&$	2.73158\times10^{-12} $&$  \text{NGC4258}    $&$                3.9\times10^7 $&$	        7.20 $&$	2.58931\times10^{-13}$\\
$\text{NGC4374(M84)} $&$	1.0\times10^9 $&$       18.4   $&$	2.59797\times10^{-12} $&$  \text{NGC4596}    $&$	        7.8\times10^7 $&$	        16.8 $&$	2.21941\times10^{-13}$\\
$\text{NGC3998}      $&$	5.6\times10^8 $&$	14.1   $&$	1.89855\times10^{-12} $&$  \text{NGC4459}    $&$	        7.0\times10^7 $&$	        16.1 $&$	2.07838\times10^{-13}$\\
$\text{NGC4486B}     $&$	6.0\times10^8 $&$	16.1   $&$	1.78147\times10^{-12} $&$  \text{NGC1023}    $&$  	        4.4\times10^7 $&$	        11.4 $&$	1.84502\times10^{-13}$\\
$\text{NGC4350}	     $&$        6.0\times10^8 $&$	16.8   $&$      1.70724\times10^{-12} $&$  \text{NGC4564}    $&$	        5.6\times10^7 $&$	        15.0 $&$	1.78463\times10^{-13}$\\
$\text{NGC4342}      $&$	3.1\times10^8 $&$	15.3   $&$	9.68551\times10^{-13} $&$  \text{NGC221(M32)}$&$  	        2.9\times10^6 $&$	        0.810$&$	1.71145\times10^{-13}$\\
$ \text{NGC3031(M81)}$&$	6.8\times10^7 $&$	3.90   $&$ 	8.33483\times10^{-13} $&$  \text{NGC821}     $&$	        8.5\times10^7 $&$	        24.1 $&$	1.68599\times10^{-13}$\\
$\text{NGC4261}	     $&$        5.2\times10^8 $&$	31.6   $&$	7.86627\times10^{-13} $&$  \text{NGC3384}    $&$  	        1.6\times10^7 $&$	        11.6 $&$	6.59348\times10^{-14}$\\
$\text{NGC4697}      $&$	1.7\times10^8 $&$	11.7   $&$	6.94569\times10^{-13} $&$  \text{NGC1068}    $&$ 	        1.5\times10^7 $&$	        15.0 $&$	4.78027\times10^{-14}$\\
$\text{CygnusA}      $&$	2.9\times10^9 $&$	240.   $&$	5.77616\times10^{-13} $&$  \text{NGC4742}    $&$	        1.4\times10^7 $&$	        15.5 $&$	4.31766\times10^{-14}$\\
$\text{NGC4291}	     $&$        3.1\times10^8 $&$	26.2   $&$	5.65604\times10^{-13} $&$  \text{NGC7332}    $&$	        1.5\times10^7 $&$   	        23.0 $&$	3.11757\times10^{-14}$\\
$\text{NGC3245}	     $&$        2.1\times10^8 $&$	20.9   $&$	4.80314\times10^{-13} $&$  \text{NGC2778}    $&$                1.4\times10^7 $&$	        22.9 $&$	2.92244\times10^{-14}$\\
$\text{NGC3379}      $&$	1.0\times10^8 $&$   	10.6   $&$	4.50969\times10^{-13} $&$  \text{NGC4945}    $&$                1.4\times10^6 $&$	        3.70 $&$	1.80875\times10^{-14}$\\
 \end{tabular}
\end{ruledtabular}
 \end{table*}
 \endgroup

\begingroup
\squeezetable
\begin{table*}
\caption{\label{tab:Table6}
Angular positions, magnifications, and time  delays of primary and secondary images due to GL by MDOs (modeled as  Schwarzschild black
holes) at centers of   many galaxies.  $\theta$,  $\mu$, and $\tau$ stand, respectively, for  angular positions, magnifications and time delays of images; subscripts
$p$ and $s$  attached to  them stand, respectively, for primary and secondary images. $\tau_s-\tau_p$ stands for the differential time
delay of the secondary image with respect to the primary image.
The time delays and differential time delays are given in {\em minutes}, whereas angular positions of  images are expressed
in {\em arcsec}. 
{\bf (a)} The first column gives the names of galaxies having MDOs  with  decreasing value of the ratio of mass to the distance (i.e., $M/D_d$). 
The ratio of the lens-source distance to the observer-source 
distance ${\cal D}=0.5$. The angular source position  $\beta=1 \mu as$.
}
\begin{ruledtabular}
  \begin{tabular}{l|cccc|ccc}  
\multicolumn{1}{c|}{MDO in}&
\multicolumn{4}{c|}{Secondary image}&
\multicolumn{3}{c}{Primary image\vspace*{-0.05in}}\\
  $\hspace*{0.3in}\text{galaxy}$  &    $\theta_{s}$            &       $\mu_{s}$     &   $\tau_{s}$     & $\tau_{s}-\tau_{p}$  &    $\theta_{p}$    &    $\mu_{p}$   &   $\tau_{p}$    \\
\hline
$\text{Milky Way}    $&$ -1.388176 $&$ -694084.2 $&$ 14.92209 $&$ 1.71\times 10^{-6} $&$ 1.388177 $&$ 694085.2 $&$ 14.92209 $\\
 $\text{NGC4486(M87)} $&$ -0.870593 $&$ -435294.8 $&$ 12859.74 $&$ 0.00226            $&$ 0.870594 $&$ 435295.8 $&$ 12859.74 $\\
 $\text{NGC4649}      $&$ -0.695869 $&$ -347933.4 $&$ 8720.112 $&$ 0.00189            $&$ 0.695870 $&$ 347934.4 $&$ 8720.110 $\\
 $\text{NGC4594}      $&$ -0.675695 $&$ -337846.2 $&$ 4806.677 $&$ 0.00107            $&$ 0.675696 $&$ 337847.2 $&$ 4806.675 $\\
 $\text{NGC3115}      $&$ -0.647562 $&$ -323779.8 $&$ 4383.655 $&$ 0.00101            $&$ 0.647563 $&$ 323780.8 $&$ 4383.654 $\\
 $\text{NGC224(M31)}  $&$ -0.612081 $&$ -306039.8 $&$ 308.1496 $&$ 0.00008            $&$ 0.612082 $&$ 306040.8 $&$ 308.1495 $\\
 $\text{IC1459}       $&$ -0.590127 $&$ -295062.6 $&$ 11035.30 $&$ 0.00278            $&$ 0.590128 $&$ 295063.6 $&$ 11035.29 $\\
 $\text{NGC5128(cenA)}$&$ -0.482112 $&$ -241055.1 $&$ 1075.303 $&$ 0.00033            $&$ 0.482113 $&$ 241056.1 $&$ 1075.302 $\\
 $\text{NGC4374(M84)} $&$ -0.470173 $&$ -235085.9 $&$ 4488.653 $&$ 0.00140            $&$ 0.470174 $&$ 235086.9 $&$ 4488.652 $\\
 $\text{NGC3998}      $&$ -0.401930 $&$ -200964.7 $&$ 2542.451 $&$ 0.00091            $&$ 0.401931 $&$ 200965.7 $&$ 2542.450 $\\
 $\text{NGC4486B}     $&$ -0.389340 $&$ -194669.5 $&$ 2730.318 $&$ 0.00101            $&$ 0.389341 $&$ 194670.5 $&$ 2730.317 $\\
 $\text{NGC4350}      $&$ -0.381143 $&$ -190570.7 $&$ 2734.506 $&$ 0.00103            $&$ 0.381144 $&$ 190571.7 $&$ 2734.505 $\\
 $\text{NGC4342}      $&$ -0.287079 $&$ -143539.0 $&$ 1441.646 $&$ 0.00071            $&$ 0.287080 $&$ 143540.0 $&$ 1441.646 $\\
 $\text{NGC3031(M81)} $&$ -0.266310 $&$ -133154.8 $&$ 317.9070 $&$ 0.00017            $&$ 0.266311 $&$ 133155.8 $&$ 317.9068 $\\
 $\text{NGC4261}      $&$ -0.258716 $&$ -129357.8 $&$ 2435.988 $&$ 0.00132            $&$ 0.258717 $&$ 129358.8 $&$ 2435.987 $\\
 $\text{NGC4697}      $&$ -0.243107 $&$ -121553.1 $&$ 799.8507 $&$ 0.00046            $&$ 0.243108 $&$ 121554.1 $&$ 799.8503 $\\
 $\text{CygnusA}      $&$ -0.221697 $&$ -110848.0 $&$ 13732.21 $&$ 0.00858            $&$ 0.221698 $&$ 110849.0 $&$ 13732.20 $\\
 $\text{NGC4291}      $&$ -0.219379 $&$ -109689.4 $&$ 1468.994 $&$ 0.00093            $&$ 0.219380 $&$ 109690.4 $&$ 1468.993 $\\
 $\text{NGC3245}      $&$ -0.202163 $&$ -101081.3 $&$ 1000.754 $&$ 0.00068            $&$ 0.202164 $&$ 101082.3 $&$ 1000.754 $\\
 $\text{NGC3379}      $&$ -0.195890 $&$ -97944.75 $&$ 477.5836 $&$ 0.00034            $&$ 0.195891 $&$ 97945.75 $&$ 477.5833 $\\
 $\text{NGC5845}      $&$ -0.194143 $&$ -97071.07 $&$ 1146.906 $&$ 0.00081            $&$ 0.194144 $&$ 97072.07 $&$ 1146.905 $\\
 $\text{NGC3377}      $&$ -0.190571 $&$ -95285.11 $&$ 478.4866 $&$ 0.00034            $&$ 0.190572 $&$ 95286.11 $&$ 478.4862 $\\
 $\text{NGC3608}      $&$ -0.183707 $&$ -91852.93 $&$ 911.4108 $&$ 0.00068            $&$ 0.183708 $&$ 91853.93 $&$ 911.4101 $\\
 $\text{NGC4473}      $&$ -0.168815 $&$ -84407.35 $&$ 530.7089 $&$ 0.00043            $&$ 0.168816 $&$ 84408.35 $&$ 530.7085 $\\
 $\text{NGC6251}      $&$ -0.152252 $&$ -76125.46 $&$ 2575.005 $&$ 0.00228            $&$ 0.152253 $&$ 76126.46 $&$ 2575.003 $\\
 $\text{NGC7052}      $&$ -0.151218 $&$ -75608.63 $&$ 1604.042 $&$ 0.00143            $&$ 0.151219 $&$ 75609.63 $&$ 1604.041 $\\
 $\text{NGC2787}      $&$ -0.149117 $&$ -74558.08 $&$ 199.4783 $&$ 0.00018            $&$ 0.149118 $&$ 74559.08 $&$ 199.4781 $\\
 $\text{NGC4258}      $&$ -0.148433 $&$ -74216.33 $&$ 189.8064 $&$ 0.00017            $&$ 0.148434 $&$ 74217.33 $&$ 189.8062 $\\
 $\text{NGC4596}      $&$ -0.137422 $&$ -68710.93 $&$ 381.5847 $&$ 0.00037            $&$ 0.137423 $&$ 68711.93 $&$ 381.5844 $\\
 $\text{NGC4459}      $&$ -0.132985 $&$ -66491.96 $&$ 343.2016 $&$ 0.00035            $&$ 0.132986 $&$ 66492.96 $&$ 343.2012 $\\
 $\text{NGC1023}      $&$ -0.125296 $&$ -62647.94 $&$ 216.5861 $&$ 0.00023            $&$ 0.125297 $&$ 62648.94 $&$ 216.5859 $\\
 $\text{NGC4564}      $&$ -0.123229 $&$ -61614.25 $&$ 275.9607 $&$ 0.00030            $&$ 0.123230 $&$ 61615.25 $&$ 275.9604 $\\
 $\text{NGC221(M32)}  $&$ -0.120676 $&$ -60337.77 $&$ 14.31073 $&$ 0.00002            $&$ 0.120677 $&$ 60338.77 $&$ 14.31072 $\\
 $\text{NGC821}       $&$ -0.119775 $&$ -59887.15 $&$ 419.6616 $&$ 0.00047            $&$ 0.119776 $&$ 59888.15 $&$ 419.6611 $\\
 $\text{NGC3384}      $&$ -0.074902 $&$ -37450.84 $&$ 81.45874 $&$ 0.00014            $&$ 0.074903 $&$ 37451.84 $&$ 81.45860 $\\
 $\text{NGC1068}      $&$ -0.063777 $&$ -31888.16 $&$ 77.15868 $&$ 0.00015            $&$ 0.063778 $&$ 31889.16 $&$ 77.15853 $\\
 $\text{NGC4742}      $&$ -0.060612 $&$ -30305.91 $&$ 72.24847 $&$ 0.00015            $&$ 0.060613 $&$ 30306.91 $&$ 72.24832 $\\
 $\text{NGC7332}      $&$ -0.051504 $&$ -25751.92 $&$ 78.21023 $&$ 0.00019            $&$ 0.051505 $&$ 25752.92 $&$ 78.21004 $\\
 $\text{NGC2778}      $&$ -0.049866 $&$ -24932.96 $&$ 73.14462 $&$ 0.00018            $&$ 0.049867 $&$ 24933.96 $&$ 73.14444 $\\
 $\text{NGC4945}      $&$ -0.039231 $&$ -19615.00 $&$ 7.424624 $&$ 0.00002            $&$ 0.039232 $&$ 19616.00 $&$ 7.424600$\\

 
 \end{tabular}
\end{ruledtabular}
 \end{table*}
 \endgroup

\begingroup
\squeezetable
\begin{table*}
\caption{\label{tab:Table7} 
 Angular positions, magnifications, and time  delays of the first and second order relativistic images on the same side as the primary image.
MDOs at centers  of many galaxies are modeled as Schwarzschild black hole lenses.  $\theta$,  $\mu$, and $\tau$ stand, respectively, for
 angular positions, magnifications, and time delays of images. Subscript $p$ stands for the primary image, whereas  $1p$, and $2p$  stand, respectively,
for the first and second order relativistic images on the same side as the primary image. Angular positions of the  images are expressed 
in $\mu as$, whereas   the time delays and differential time delays are given in {\em minutes}. 
Other inputs are the same as {\bf (a)} of Table V.
}
\begin{ruledtabular}
  \begin{tabular}{l|cccc|cccc}  
\multicolumn{1}{c|}{$\text{MDO in}$}&
\multicolumn{4}{c|}{Second order  relativistic image}&
\multicolumn{4}{c} {First order  relativistic image \vspace*{-0.05in}}\\
 $\hspace*{0.3in}\text{galaxy}$  &  $\theta_{2p}$ &    $\mu_{2p}$ & $\tau_{2p}$  & $\tau_{2p}-\tau_{1p}$ &  $\theta_{1p}$ & $\mu_{1p}$ & $\tau_{1p}$  & $\tau_{1p}-\tau_{p}$   \\
\hline
$ \text{Milky Way}    $&$ 24.272396 $&$  1.34\times10^{-14} $&$ 48.02749 $&$ 9.66664 $&$ 24.302833 $&$       7.21\times10^{-12} $&$ 38.36085 $&$ 23.43876 $\\  
$ \text{NGC4486(M87)} $&$ 9.5467556 $&$  2.07\times10^{-15} $&$ 40830.26 $&$ 8033.22 $&$ 9.5587269 $&$       1.12\times10^{-12} $&$ 32797.05 $&$ 19937.31 $\\  
$ \text{NGC4649}      $&$ 6.0993161 $&$  8.43\times10^{-16} $&$ 27514.08 $&$ 5355.48 $&$ 6.1069644 $&$       4.55\times10^{-13} $&$ 22158.61 $&$ 13438.50 $\\  
$ \text{NGC4594}      $&$ 5.7507837 $&$  7.49\times10^{-16} $&$ 15153.98 $&$ 2945.51 $&$ 5.7579950 $&$       4.05\times10^{-13} $&$ 12208.46 $&$ 7401.788 $\\  
$ \text{NGC3115}      $&$ 5.2818819 $&$  6.32\times10^{-16} $&$ 13804.24 $&$ 2677.74 $&$ 5.2885053 $&$       3.41\times10^{-13} $&$ 11126.50 $&$ 6742.847 $\\  
$ \text{NGC224(M31)}  $&$ 4.7189445 $&$  5.05\times10^{-16} $&$ 968.8844 $&$ 187.442 $&$ 4.7248619 $&$       2.73\times10^{-13} $&$ 781.4426 $&$ 473.2931 $\\  
$ \text{IC1459}       $&$ 4.3864944 $&$  4.36\times10^{-16} $&$ 34662.92 $&$ 6694.35 $&$ 4.3919950 $&$       2.36\times10^{-13} $&$ 27968.57 $&$ 16933.28 $\\  
$ \text{NGC5128(cenA)}$&$ 2.9276717 $&$  1.94\times10^{-16} $&$ 3359.469 $&$ 642.657 $&$ 2.9313429 $&$       1.05\times10^{-13} $&$ 2716.812 $&$ 1641.509 $\\  
$ \text{NGC4374(M84)} $&$ 2.7844704 $&$  1.76\times10^{-16} $&$ 14014.24 $&$ 2677.74 $&$ 2.7879620 $&$       9.49\times10^{-14} $&$ 11336.50 $&$ 6847.846 $\\  
$ \text{NGC3998}      $&$ 2.0348357 $&$  9.38\times10^{-17} $&$ 7905.583 $&$ 1499.53 $&$ 2.0373873 $&$       5.07\times10^{-14} $&$ 6406.049 $&$ 3863.599 $\\  
$ \text{NGC4486B}     $&$ 1.9093511 $&$  8.26\times10^{-17} $&$ 8482.795 $&$ 1606.64 $&$ 1.9117454 $&$       4.46\times10^{-14} $&$ 6876.151 $&$ 4145.834 $\\  
$ \text{NGC4350}      $&$ 1.8297948 $&$  7.59\times10^{-17} $&$ 8491.171 $&$ 1606.64 $&$ 1.8320893 $&$       4.10\times10^{-14} $&$ 6884.527 $&$ 4150.022 $\\  
$ \text{NGC4342}      $&$ 1.0380797 $&$  2.44\times10^{-17} $&$ 4444.741 $&$ 830.099 $&$ 1.0393814 $&$       1.32\times10^{-14} $&$ 3614.642 $&$ 2172.996 $\\  
$ \text{NGC3031(M81)} $&$ 0.8933152 $&$  1.81\times10^{-17} $&$ 978.3253 $&$ 182.086 $&$ 0.8944354 $&$       9.77\times10^{-15} $&$ 796.2390 $&$ 478.3322 $\\  
$ \text{NGC4261}      $&$ 0.8430953 $&$  1.61\times10^{-17} $&$ 7491.180 $&$ 1392.42 $&$ 0.8441526 $&$       8.70\times10^{-15} $&$ 6098.755 $&$ 3662.769 $\\  
$ \text{NGC4697}      $&$ 0.7444293 $&$  1.26\times10^{-17} $&$ 2455.980 $&$ 455.216 $&$ 0.7453628 $&$       6.78\times10^{-15} $&$ 2000.764 $&$ 1200.914 $\\  
$ \text{CygnusA}      $&$ 0.6190806 $&$  8.69\times10^{-18} $&$ 42071.51 $&$ 7765.44 $&$ 0.6198569 $&$       4.69\times10^{-15} $&$ 34306.07 $&$ 20573.87 $\\  
$ \text{NGC4291}      $&$ 0.6062068 $&$  8.33\times10^{-18} $&$ 4499.436 $&$ 830.099 $&$ 0.6069670 $&$       4.50\times10^{-15} $&$ 3669.337 $&$ 2200.344 $\\  
$ \text{NGC3245}      $&$ 0.5147940 $&$  6.01\times10^{-18} $&$ 3059.264 $&$ 562.325 $&$ 0.5154395 $&$       3.24\times10^{-15} $&$ 2496.939 $&$ 1496.185 $\\  
$ \text{NGC3379}      $&$ 0.4833420 $&$  5.29\times10^{-18} $&$ 1458.860 $&$ 267.774 $&$ 0.4839481 $&$       2.86\times10^{-15} $&$ 1191.086 $&$ 713.5030 $\\  
$ \text{NGC5845}      $&$ 0.4747576 $&$  5.11\times10^{-18} $&$ 3502.675 $&$ 642.657 $&$ 0.4753529 $&$       2.76\times10^{-15} $&$ 2860.018 $&$ 1713.113 $\\  
$ \text{NGC3377}      $&$ 0.4574487 $&$  4.74\times10^{-18} $&$ 1460.666 $&$ 267.774 $&$ 0.4580223 $&$       2.56\times10^{-15} $&$ 1192.892 $&$ 714.4060 $\\  
$ \text{NGC3608}      $&$ 0.4250877 $&$  4.09\times10^{-18} $&$ 2779.838 $&$ 508.770 $&$ 0.4256208 $&$       2.21\times10^{-15} $&$ 2271.068 $&$ 1359.658 $\\  
$ \text{NGC4473}      $&$ 0.3589661 $&$  2.92\times10^{-18} $&$ 1615.480 $&$ 294.551 $&$ 0.3594163 $&$       1.58\times10^{-15} $&$ 1320.929 $&$ 790.2203 $\\  
$ \text{NGC6251}      $&$ 0.2919802 $&$  1.93\times10^{-18} $&$ 7819.582 $&$ 1419.20 $&$ 0.2923463 $&$       1.04\times10^{-15} $&$ 6400.381 $&$ 3825.378 $\\  
$ \text{NGC7052}      $&$ 0.2880290 $&$  1.88\times10^{-18} $&$ 4870.271 $&$ 883.654 $&$ 0.2883902 $&$       1.02\times10^{-15} $&$ 3986.617 $&$ 2382.577 $\\  
$ \text{NGC2787}      $&$ 0.2800806 $&$  1.78\times10^{-18} $&$ 605.4707 $&$ 109.787 $&$ 0.2804318 $&$       9.60\times10^{-16} $&$ 495.6833 $&$ 296.2053 $\\  
$ \text{NGC4258}      $&$ 0.2775189 $&$  1.75\times10^{-18} $&$ 576.0530 $&$ 104.432 $&$ 0.2778669 $&$       9.43\times10^{-16} $&$ 471.6212 $&$ 281.8150 $\\  
$ \text{NGC4596}      $&$ 0.2378733 $&$  1.28\times10^{-18} $&$ 1156.050 $&$ 208.864 $&$ 0.2381716 $&$       6.93\times10^{-16} $&$ 947.1863 $&$ 565.6019 $\\  
$ \text{NGC4459}      $&$ 0.2227576 $&$  1.12\times10^{-18} $&$ 1038.988 $&$ 187.442 $&$ 0.2230370 $&$       6.07\times10^{-16} $&$ 851.5464 $&$ 508.3452 $\\  
$ \text{NGC1023}      $&$ 0.1977463 $&$  8.86\times10^{-19} $&$ 654.7971 $&$ 117.821 $&$ 0.1979942 $&$       4.79\times10^{-16} $&$ 536.9766 $&$ 320.3907 $\\  
$ \text{NGC4564}      $&$ 0.1912746 $&$  8.29\times10^{-19} $&$ 833.9894 $&$ 149.953 $&$ 0.1915144 $&$       4.48\times10^{-16} $&$ 684.0360 $&$ 408.0756 $\\  
$ \text{NGC221(M32)}  $&$ 0.1834313 $&$  7.62\times10^{-19} $&$ 43.22856 $&$ 7.76544 $&$ 0.1836613 $&$       4.12\times10^{-16} $&$ 35.46312 $&$ 21.15240 $\\  
$ \text{NGC821}       $&$ 0.1807017 $&$  7.40\times10^{-19} $&$ 1267.462 $&$ 227.608 $&$ 0.1809283 $&$       4.00\times10^{-16} $&$ 1039.854 $&$ 620.1931 $\\  
$ \text{NGC3384}      $&$ 0.0706679 $&$  1.13\times10^{-19} $&$ 243.5083 $&$ 42.8438 $&$ 0.0707566 $&$       6.11\times10^{-17} $&$ 200.6645 $&$ 119.2059 $\\  
$ \text{NGC1068}      $&$ 0.0512343 $&$  5.95\times10^{-20} $&$ 229.8712 $&$ 40.1661 $&$ 0.0512985 $&$       3.21\times10^{-17} $&$ 189.7051 $&$ 112.5466 $\\  
$ \text{NGC4742}      $&$ 0.0462761 $&$  4.85\times10^{-20} $&$ 215.0139 $&$ 37.4883 $&$ 0.0463341 $&$       2.62\times10^{-17} $&$ 177.5255 $&$ 105.2772 $\\  
$ \text{NGC7332}      $&$ 0.0334136 $&$  2.53\times10^{-20} $&$ 231.9743 $&$ 40.1661 $&$ 0.0334555 $&$       1.37\times10^{-17} $&$ 191.8082 $&$ 113.5982 $\\  
$ \text{NGC2778}      $&$ 0.0313223 $&$  2.22\times10^{-20} $&$ 216.8061 $&$ 37.4883 $&$ 0.0313615 $&$       1.20\times10^{-17} $&$ 179.3178 $&$ 106.1734 $\\  
$ \text{NGC4945}      $&$ 0.0193859 $&$  8.52\times10^{-21} $&$ 21.90093 $&$ 3.74883 $&$ 0.0194102 $&$       4.60\times10^{-18} $&$ 18.15210 $&$ 10.72750 $\\
 \end{tabular}
\end{ruledtabular}
 \end{table*}
 \endgroup
\begingroup
\squeezetable
\begin{table*}
\caption{\label{tab:Table5}  Comparison of Bozza and Mancini's approximate analytical \cite{BozMan04} and our  numerical results. 
Mass and distance of MDOs are given in  the units of {\em solar mass} and {\em Mpc}, respectively. $M/D_d$  in the third column is dimensionless ($M\equiv MG/c^2$).
Differential time delays $\tau_{2}-\tau_{1}$ are expressed in {\em minutes}. 
$\tau_{2}$ and $\tau_{1}$ stand, respectively, for time delays of the first and  second order relativistic images for the angular source position $\beta=0$.
The ratio of
source-lens to source-observer distances  ${\cal D}=0.5$. The ratio of mass of an MDO to the differential time delay $M/(\tau_{2}-\tau_{1})$
is expressed in terms of {\em solar mass/minute}. Masses and distances  of MDOs in this table are taken the same as  in \cite{BozMan04}.
       } 
\begin{ruledtabular}
  \begin{tabular}{l|c|c|c|lcc|lr}  
\multicolumn{1}{c|}{MDO in}&
\multicolumn{1}{c|}{Mass}&
\multicolumn{1}{c|} {Distance}&
\multicolumn{1}{c|} {$M/D_d$}&
\multicolumn{3}{c|} {Differential time delay $(\tau_{2}-\tau_{1})$}&
\multicolumn{2}{c} {$M/(\tau_{2}-\tau_{1})$}\\
$\hspace*{0.4in}\text{galaxy}$  &  $M$ &    $D_d $  & $ $ &  $\text{Analytical}$ & $\text{Numerical}$ & $\text{\hspace*{0.2in} \% difference}$  &   $\text{Analytical\hspace*{0.2in}}$ &   $\text{Numerical}$ \\

\hline

 $\text{Milky Way}      $&$ 2.8\times 10^6 $&$ 0.0085  $&$    1.57\times10^{-11} $&$ 6      $&$ 7.4977   $&$ 20.0  $&$ 466666.666666667 \hspace*{0.2in}  $&$ 373449.374476538   $\\
 $\text{NGC3115}        $&$ 2.0\times 10^9 $&$ 8.4     $&$    1.14\times10^{-11} $&$ 5430   $&$ 5355.5   $&$ -1.4  $&$ 368324.125230203 \hspace*{0.2in} $&$ 373449.374476535   $\\
 $\text{NGC4486(M87)}   $&$ 3.3\times 10^9 $&$ 15.3    $&$    1.03\times10^{-11} $&$ 8958   $&$ 8836.5   $&$ -1.4  $&$ 368385.800401875 \hspace*{0.2in} $&$ 373449.374476534   $\\
 $\text{NGC4594}        $&$ 1.0\times 10^9 $&$ 9.2     $&$    5.20\times10^{-12} $&$ 2712   $&$ 2677.7   $&$ -1.3  $&$ 368731.563421829 \hspace*{0.2in} $&$ 373449.374476530   $\\
 $\text{NGC4374 (M84)}  $&$ 1.4\times 10^9 $&$ 15.3    $&$    4.37\times10^{-12} $&$ 3798   $&$ 3748.8   $&$ -1.3  $&$ 368615.060558189 \hspace*{0.2in} $&$ 373449.374476529   $\\
 $\text{NGC224(M31)}    $&$ 3.0\times 10^7 $&$ 0.7     $&$    2.05\times10^{-12} $&$ 84     $&$ 80.332   $&$ -4.6  $&$ 357142.857142857 \hspace*{0.2in} $&$ 373449.374476527   $\\
 $\text{NGC4486B(M104)} $&$ 5.7\times 10^8 $&$ 15.3    $&$    1.78\times10^{-12} $&$ 1548   $&$ 1526.3   $&$ -1.4  $&$ 368217.054263566 \hspace*{0.2in} $&$ 373449.374476527   $\\
 $\text{NGC4342(IC3256)}$&$ 3.0\times 10^8 $&$ 15.3    $&$    9.37\times10^{-13} $&$ 816    $&$ 803.32   $&$ -1.6  $&$ 367647.058823529 \hspace*{0.2in} $&$ 373449.374476527   $\\
 $\text{NGC3377}        $&$ 1.8\times 10^8 $&$ 9.9     $&$    8.69\times10^{-13} $&$ 486    $&$ 481.99   $&$ -0.8  $&$ 370370.370370370 \hspace*{0.2in} $&$ 373449.374476527   $\\
 $\text{NGC4261}        $&$ 4.5\times 10^8 $&$ 27.4    $&$    7.85\times10^{-13} $&$ 1224   $&$ 1205.0   $&$ -1.6  $&$ 367647.058823529 \hspace*{0.2in} $&$ 373449.374476526   $\\
 $\text{NGC7052}        $&$ 3.3\times 10^8 $&$ 58.7    $&$    2.69\times10^{-13} $&$ 894    $&$ 883.65   $&$ -1.2  $&$ 369127.516778523 \hspace*{0.2in} $&$ 373449.374476526   $\\
 $\text{NGC0221(M32)}   $&$ 3.4\times 10^6 $&$ 0.7     $&$    2.32\times10^{-13} $&$ 12     $&$ 9.1043   $&$ -31.8 $&$ 283333.333333333 \hspace*{0.2in} $&$ 373449.374476526   $\\

 \end{tabular}
\end{ruledtabular}
 \end{table*}
 \endgroup

\end{document}